\documentclass[lettersize,journal]{IEEEtran}
\IEEEoverridecommandlockouts
\usepackage{amsmath}
\DeclareMathOperator{\Tr}{Tr}
\usepackage{cite}
\usepackage{amsmath,amssymb,amsfonts}
\usepackage{algorithm}
\usepackage{algorithmicx}
\usepackage{algpseudocode}  % 提供 \Comment、\For、\Require 等
\usepackage{graphicx}
\usepackage{textcomp}
\usepackage{xcolor}
\usepackage{stfloats}
\usepackage{url}
\usepackage{subfig}
\usepackage{graphicx}
\usepackage{bm}
\usepackage{booktabs}
\usepackage{cite}
\usepackage{color}
\usepackage{makecell}
\usepackage{upgreek}
\usepackage{verbatim}
\def\BibTeX{{\rm B\kern-.05em{\sc i\kern-.025em b}\kern-.08em
		T\kern-.1667em\lower.7ex\hbox{E}\kern-.125emX}}
\usepackage{balance}
\usepackage{multirow}
\usepackage{url}
\usepackage{color}
\usepackage{nomencl}
\usepackage{float}
\usepackage{booktabs}
\def\BibTeX{{\rm B\kern-.05em{\sc i\kern-.025em b}\kern-.08em
    T\kern-.1667em\lower.7ex\hbox{E}\kern-.125emX}}
\begin{document}

\title{Synesthesia of Machines (SoM)-Aided Online FDD Precoding via Heterogeneous Multi-Modal Sensing: A Vertical Federated Learning Approach}
\author{Haotian Zhang,~\IEEEmembership{Graduate Student Member,~IEEE,} Shijian Gao,~\IEEEmembership{Member,~IEEE,}  Weibo Wen,~\IEEEmembership{Graduate Student Member,~IEEE,}  Xiang Cheng,~\IEEEmembership{Fellow,~IEEE,}  Liuqing Yang,~\IEEEmembership{Fellow,~IEEE} 

	\thanks{
    An earlier version of this paper was presented in part at the 2025 IEEE International Conference on Communications (ICC) \cite{ICC}, Montreal, Canada. This work was supported in part by the National Natural Science Foundation of China under Grant 62125101, Grant 62341101, Grant 62401488, and Grant U23A20339, in part by the New Cornerstone Science
    Foundation through the Xplorer Prize, in part by the Guangdong Provincial Key Laboratory of Future Networks of Intelligence, The Chinese University of Hong Kong, Shenzhen, under Grant No. 2022B1212010001-OF08, and in part by Guangzhou Municipal Project under Grant 2023A03J0011 and Grant 2024D03J0008, and Guangdong Provincial Project under Grant 2023ZDZX1037 and Grant 2023ZT10X009.

	Haotian Zhang, Weibo Wen, and Xiang Cheng are with the State Key Laboratory of Photonics and Communications, School of Electronics, Peking University, Beijing 100871, P. R. China (e-mail: haotianzhang@stu.pku.edu.cn; weber@stu.pku.edu.cn; xiangcheng@pku.edu.cn).

    Shijian Gao is with the Internet of Things Thrust, The Hong Kong University of Science and Technology (Guangzhou), Guangzhou 511400, P. R. China (e-mail: shijiangao@hkust-gz.edu.cn).  He is also with Guangdong Provincial Key Laboratory of Future Networks of Intelligence, The Chinese University of Hong Kong, Shenzhen 518172, P. R. China.

    Liuqing Yang is with the Internet of Things Thrust and Intelligent Transportation Thrust, The Hong Kong University of Science and Technology (Guangzhou), Guangzhou 510000, P. R. China, and the Department of Electronic and Computer Engineering, The Hong Kong University of Science and Technology, Hong Kong SAR, P. R. China (email: lqyang@ust.hk).

}}

\markboth{Journal of \LaTeX\ Class Files,~Vol.~18, No.~9, September~2020}%
{Synesthesia of Machines (SoM)-Aided Online FDD Precoding via Heterogeneous Multi-Modal Sensing: A Vertical Federated Learning Approach}

\maketitle

\begin{abstract}
This paper investigates a heterogeneous multi-vehicle, multi-modal sensing (H-MVMM) aided online precoding problem. The proposed H-MVMM scheme utilizes a vertical federated learning (VFL) framework to minimize pilot sequence length and optimize the sum rate.  This offers a promising solution for reducing latency in frequency division duplexing systems. To achieve this, three preprocessing modules are designed to transform raw sensory data into informative representations relevant to precoding. The approach effectively addresses local data heterogeneity arising from diverse on-board sensor configurations through a well-structured VFL training procedure. Additionally, a label-free online model updating strategy is introduced, enabling the H-MVMM scheme to adapt its weights flexibly. This strategy features a pseudo downlink channel state information label simulator (PCSI-Simulator), which is trained using a semi-supervised learning (SSL) approach alongside an online loss function. Numerical results show that the proposed method can closely approximate the performance of traditional optimization techniques with perfect channel state information, achieving a significant 90.6\% reduction in pilot sequence length.
\end{abstract}

% \vspace{-1.0em}
\section{Introduction}
Massive multiple-input multiple-output (mMIMO) is a key  booster for 6G wireless networks \cite{mMIMO1,mMIMO2}. To mitigate inter-user interference and enhance throughput, it is essential to carefully consider the precoding design at the road side unit (RSU). In frequency division duplexing (FDD) systems, the precoding design necessitates downlink (DL) channel estimation and uplink (UL) feedback to acquire DL channel state information (CSI). This process incurs long latency under mMIMO, eroding the potential benefits of FDD systems. 
In latency-sensitive applications like vehicular communication networks (VCN) \cite{VCN, FanV2V, huangscatter}, optimizing precoding while reducing pilot overhead has emerged as a critical and challenging problem.

Many studies have focused on precoding optimization. Conventional approaches like zero-forcing (ZF) \cite{ZF} and weighted minimum mean square error (WMMSE) method \cite{WMMSE} can offer outstanding performance under perfect CSI but suffer from high pilot overhead and excessive complexity. Such issues are also faced by many centralized optimization schemes \cite{opt-1,opt-2}.  
Towards pilot overhead reduction, compressed sensing (CS) based methods leveraging channel sparsity have been extensively studied \cite{CS1,CS2}. However, they primarily rely on sparsity as prior information, and their computational complexity is high since solving CS problems typically involves cumbersome iterations. 
Recent works have introduced deep learning for FDD precoding optimization \cite{Weiyu-MISO,Weiyu-MIMO}. Yet, we observe that they still require a long DL pilot sequence to approach the performance  comparable to that of ZF or WMMSE schemes.

With the increased level of intelligence in VCN, many vehicles have been equipped with various sensing devices, such as Light Detection and Ranging (LiDAR) and red-green-blue (RGB) cameras, providing vehicles with abundant environmental information \cite{VCN}. Recently, Cheng {\em et al.} propose the Synesthesia of Machines (SoM) that offers a new perspective from the multi-modal sensing domain to decrease pilot overhead \cite{SoM}. {\color{black}To date, many studies have explored utilizing multi-modal sensing to partially replace the role of pilots and enhance wireless communication system performance \cite{b3,FL_beam_jiahui,MMFF,CLN,gff-link,gff-bsselection,MM_beam_generative}. However, limited efforts have been devoted to leveraging multi-modal sensing for precoding design. Moreover, most existing works are limited to utilizing multi-modal sensors deployed at RSUs. The sensing range of RSU-mounted sensors is limited, restricting their optimization to line-of-sight (LoS) links. To address this, vehicle-mounted sensors can be utilized to capture environmental features near the users.
So far, only a few studies have recognized the unique role of vehicle-mounted sensors and utilized them to optimize transceiver design. In \cite{b3,FL_beam_jiahui}, vehicle-mounted sensors are used to guide optimal beam selection at the user side, which cannot be applied to multi-user precoding design at BS.
In this work, we explore the use of vehicle-mounted multi-modal sensors to optimize precoding performance.}

Considering the significant workforce and privacy issues associated with dataset collection process in traditional centralized learning (CL), we adopt a federated learning (FL) framework. Note that vehicles differ in their levels of intelligence, resulting in varied sensor configurations and, consequently, potential data heterogeneity. This presents challenges to FL training by impeding the model convergence. Existing FL-based works \cite{beamformingFL2, beamformingFL3} typically adopt homogeneous local datasets, {\color{black} where the data categories and structures across different datasets are identical, differing only in data volume. Little attention has been paid to this underexplored data heterogeneity issues.}
Furthermore, precoding design requires analyzing the correlation among multi-user CSI. Such unique characteristic entails that all involved vehicles collaboratively train a precoding model, with each local {\color{black}neural network (NN)} serving as a distinct component. This differs from most existing FL-based precoding works where all users share and update a unified global model \cite{b3,beamformingFL2,b9}. In summary, designing a customized FL framework capable of accommodating the heterogeneity of local datasets and the unique characteristic of precoding design presents a significant challenge.

Traditional DL-based schemes typically designed for a specific user number or in static channel conditions \cite{Weiyu-MISO}. Due to the varying nature of VCN, these methods require pre-training and storing numerous models for various system configurations and channel characteristics, which is highly inconvenient and inefficient. {\color{black}For instance, a model pre-trained under low vehicle density may fail to maintain its performance in dense traffic scenarios, as the interference pattern and channel availability change drastically. Additionally, dynamic channel conditions, like sudden signal blockage from large trucks or varying weather impacts, further challenge models trained on static or historical channel samples.} To address these limitations, this work aims to enable the precoding NN to update flexibly online. Existing FL-based precoding works \cite{beamformingFL2, beamformingFL3} cannot adapt to the varying system configurations or channel conditions during online inference stage, which limits their applicability in VCNs. A major challenge for online model updating is the prohibitive overhead of acquiring real-time CSI labels through channel estimation in FDD systems. Hence, the design of label-free online NN updating mechanism holds significant practical implications. Currently, some researchers have attempted to use real-time measurements to update NN without the need for true labels or reduce the dataset collection overhead \cite{onlinework-1,onlinework-2,onlinework-3,FDD_data_augment}. In \cite{onlinework-1,onlinework-2}, online training frameworks are proposed to realize label-free channel estimation using real-time received pilots. However, in the low signal-to-noise-ratio (SNR) regime, those noise-corrupted signals induce complex signal-channel mappings that hinder NN's convergence. Jha {\em et al.} \cite{onlinework-3} proposed an online Bayesian NN for channel estimation. However, its dependence on fixed hidden Markov model priors reduces its adaptability to non-stationary channels. Up to now, label-free online updating for precoding NNs in FDD systems remains an open issue.

In this paper, we propose a heterogeneous multi-vehicle, multi-modal (H-MVMM) sensing aided precoding method guided by SoM. By leveraging multi-view environmental features, the H-MVMM scheme effectively reduces the pilot overhead while ensuring a high system sum rate. Specifically, we adopt the vertical FL (VFL) framework to implement H-MVMM since VFL is tailored for situations where users have distinct feature spaces. To bridge the substantial gap between multi-modal sensing and precoding, we customize three unique data preprocessing modules to transform raw sensory data into precoding-related representations, which have been proven to effectively enhance the functionality of pilot transmission. To mitigate the effects of data heterogeneity on VFL training, we design a customized loss function that can accelerate local model convergence. The H-MVMM scheme is further endowed with the ability to perform online updates without requiring true CSI labels. Specifically, we design a pseudo DL CSI label simulator (PCSI-Simulator) to provide RSU with pseudo labels for online model updating. The PCSI-Simulator consists of three NN components: the Main CSI Recovering NN (Main-CSI-Net), the Residual CSI Fitting NN (Resi-CSI-Net), and the Residual Codeword Selection NN (Resi-CW-Net). Thanks to the proposed label-free training methodology, none of these components require true labels for training. Specifically, we leverage an online loss function capable of measuring the error of the generated pseudo DL CSI. For the label-free training of Resi-CW-Net, we propose a semi-supervised learning (SSL) approach that uses a small amount of labeled data to train a Codeword-Teacher NN, which then generates pseudo codeword index labels for Resi-CW-Net.  
Our key contributions are summarized as follows:

\begin{itemize}
\item[ $\bullet$]
   
    Heterogeneous multi-vehicle multi-modal sensing is utilized to optimize precoding performance while reducing pilot overhead. 
    The H-MVMM scheme is built upon VFL framework, where the training procedure is carefully designed to mitigate the adverse effects of local dataset heterogeneity on VFL training. 

\end{itemize}

\begin{itemize}
    \item[ $\bullet$]
    A label-free online model updating strategy for H-MVMM is proposed, where the PCSI-Simulator is developed as a pseudo CSI label provider. The label-free training of PCSI-Simulator is achieved via an online loss function as well as an SSL-based pseudo codeword index label generation approach.
 
\end{itemize}

\begin{itemize}
    \item[ $\bullet$]
     Extensive numerical results validate the proposed schemes, confirming that the H-MVMM scheme can approach the sum rate performance of the WMMSE method with perfect CSI while reducing pilot overhead by over 90\%. Additionally, a convergence analysis of the PCSI-Simulator is provided.
\end{itemize}

\textit{Notations:} In this paper, we use a capital boldface letter $\mathbf{A}$ to denote a matrix, a lowercase boldface letter $\mathbf{a}$ to denote a vector, and a lowercase letter $a$ to denote a scalar, respectively. $\mathbf{A}[:,m]$ represents the $m$-th column of $\mathbf{A}$. $\mathbf{A}^{-1}$, $\mathbf{A}^{\rm{H}}$, and $\mathbf{A}^{\dagger}$ represent the inverse,  Hermitian, and pseudo-inverse  of $\mathbf{A}$, respectively.  $\lvert \cdot \rvert$, $\Vert \cdot \Vert_{\rm{F}}$, $\Vert \cdot \Vert_{1}$, and $\Vert \cdot \Vert_{2}$ stand for the modulus of a complex number, the Frobenius norm of a matrix, $l_1$-norm and $l_2$-norm of a vector, respectively. The notations $\mathbb{E[\cdot]}$, $\Tr(\cdot)$, and $\oplus$ represent the expectation, matrix trace, the concatenation operation, respectively. Given vectors $\mathbf{a} \in \mathbb{C} ^{1\times M}$ and $\mathbf{b} \in \mathbb{C} ^{1 \times N}$, $\mathbf{a} \oplus \mathbf{b} = [\mathbf{a},\mathbf{b}] \in \mathbb{C} ^{1\times (M+N)}$. The ${\text{top-}K(\mathbf{a})}$ operator returns the $K$-th largest value in vector $\mathbf{a}$. The operator $\eta$ denotes the isomorphic mapping that transforms a complex-valued vector $\mathbf{a}$ into its real-valued counterpart $\tilde{\mathbf{a}}$: $\eta: \mathbf{a} \mapsto \tilde{\mathbf{a}}=\text{Re}(\mathbf{a})\oplus\text{Im}(\mathbf{\mathbf{a}})$. 

% \vspace{-0.2em}
\section{System Model and Problem Formulation}
\label{section ii}
We consider an FDD multi-user MIMO system where an RSU is equipped with an $ N_{\rm v} \times N_{\rm h}$ uniform planar array, serving $K$ single-antenna users. $N_{\rm v}$ and $N_{\rm h}$ denote the number of vertical and horizontal antennas, with $N=N_{\rm v} N_{\rm h}$. Suppose the RSU employs linear precoding to transmit the data symbols $s_k$ for each user. The transmitted signal can be expressed as:
% \vspace{-1em}
\begin{equation}
\label{V}
% \vspace{-0.5em}
\mathbf{x} = \sum^K_{k=1}\mathbf{v}_k s_k = \mathbf{V}\mathbf{s},
% \vspace{-0.2em}
\end{equation}
where $\mathbf{v}_k \in \mathbb{C}^{N \times 1}$ represents the $k$-th {\color{black}column} of the precoding matrix $\mathbf{V} \in \mathbb{C}^{N \times K}$. The precoding matrix $\mathbf{V}$ satisfies $\Tr(\mathbf{V}\mathbf{V}^{\rm H}) \leq P $. The power of data symbols $\mathbf{s} \in \mathbb{C}^{K\times1}$ is also normalized so that $\mathbb{E}[\mathbf{s}\mathbf{s}^{\rm H}]=\mathbf{I}_K$. 

Let $\mathbf{h}_k \in \mathbb{C}^{N \times 1}$ stand for the channel between RSU and the $k$-th user. The received signal at the $k$-th user in DL data transmission stage can be expressed as $y_k = \mathbf{h}_k^{\rm H}\mathbf{v}_k s_k + \sum_{i \neq k}\mathbf{h}_k^{\rm H}\mathbf{v}_i s_i + n_k$, where $n_k \sim \mathcal{CN}(0,\sigma^2)$ is the additive white Gaussian noise. Therefore, the achievable rate for user-$k$ is computed as:
\begin{equation}
\label{Y}
 % \vspace{-0.5em}
R_k = \log_2(1+\frac{|\mathbf{h}_k^{\rm H}\mathbf{v}_k|^2}{\sum_{i \neq k}|\mathbf{h}_k^{\rm H}\mathbf{v}_i|^2+\sigma^2}).
 % \vspace{-0.5em}
\end{equation}
\begin{figure*}[!t]
	\centering
	\includegraphics[width=1\linewidth]{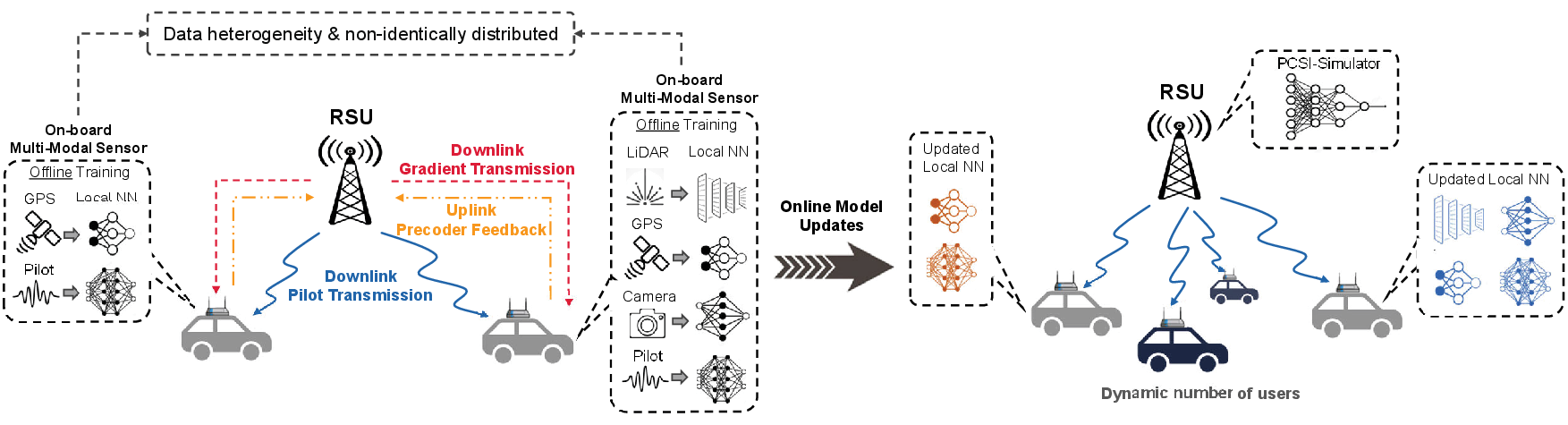}
	\caption{An illustration of the system model.
	\label{wholesystem}}
%     \vspace{-1em}
\end{figure*}
To design $\mathbf{V}$, RSU needs instantaneous DL CSI, typically acquired through estimation and feedback. 
In our scheme, similar to \cite{Weiyu-MISO}, we propose to bypass explicit user-end channel estimation and directly design the precoder based on received signals.
In DL training stage, RSU broadcasts pilot sequence $\mathbf{X} \in \mathbb{C}^{N\times L_{\rm p}}$ of length $L_{\rm P}$ and the received signals $\mathbf{y}_k \in \mathbb{C}^{1 \times L_{\rm p}}$ at the $k$-th user can be modeled as:
\begin{equation}
\label{Y}
       \mathbf{y}_k = \mathbf{h}_k^{\rm H} \mathbf{X} + \mathbf{z}_k,\\
\end{equation}
where $\mathbf{z}_k  \sim \mathcal{CN}(\mathbf{0},\sigma^2\mathbf{I}_{L_{\rm P}}) \in \mathbb{C}^{1\times L_{\rm p}}$ is the additive white Gaussian noise and the pilots transmitted at the $l$-th time instant are subject to the transmit power budget $P$ as $\Vert \mathbf{x}_l \Vert_2^2 \leq P$. After the $k$-th user obtains its precoding vector $\mathbf{v}_k$ using its local NN, it subsequently feeds back $\mathbf{v}_k$ to RSU in form of $Q$ bits as:  $\mathbf{b}_k = \mathcal{F}(\mathbf{v}_k)$, where function $\mathcal{F}: \mathbb{C}^{N\times1} \rightarrow \{\pm1\}^Q$ represents a certain quantization scheme.\footnote{This work centers on improving FDD precoding through heterogeneous multi-modal sensing. To facilitate the feedback of the precoding vector $\mathbf{v}_k$, a $B$-bit uniform quantization strategy is employed, in which both the real and imaginary components of a complex number are represented using $B$ bits.}

Given the above established model, the task of maximizing the achievable rate can be formulated as:
\vspace{-0.4em}
\begin{subequations}
\begin{alignat}{4}
\mathop{\max}\limits_{\mathbf{\Theta}_1,\cdots,\mathbf{\Theta}_K}~ & \sum_{k=1}^K ~ R_k&{}& 
\label{obj}\\
\mbox{s.t.}\quad
% & \mathbf{b}_k = \mathcal{F}(\mathbf{v}_k) = \mathcal{F}\Big(\eta^{-1}\big(\mathcal{G}_k(\eta(\mathbf{y}_k), \mathbb{M}_k)\big)\Big) ,  ~~\forall k \\
&\mathbf{V} = [\mathbf{v}_1,\cdots,\mathbf{v}_K],\\
% &\enspace \; \, = [\mathcal{D}(\mathbf{b}_1),\cdots,\mathcal{D}(\mathbf{b}_K)]\notag \\
&\enspace \; \, = [\mathcal{D}\big(\mathcal{F}(\hat{\mathbf{v}}_1)\big),\cdots,\mathcal{D}\big(\mathcal{F}(\hat{\mathbf{v}}_K)\big)]\notag \\
& \Tr(\mathbf{V}\mathbf{V}^{\rm H}) \leq P , 
% &\Vert \mathbf{x}_l \Vert_2^2 \leq P,  ~~ \forall l,
% % &L_{\rm p} \leq L_{\rm T},
\end{alignat}
\end{subequations}
where $\hat{\mathbf{v}}_k=\eta^{-1}\big(\mathcal{G}_k(\eta(\mathbf{y}_k), \mathbb{M}_k)\big)$ represents the original output of the $k$-th vehicle's local model $\mathcal{G}_{k}(\cdot)$, which is parameterized by $\mathbf{\Theta}_k, k=1,\cdots,K$. ${\mathbf{v}}_k=\mathcal{D}\big(\mathcal{F}(\hat{\mathbf{v}}_K)\big)$ represents the $k$-th vehicle's precoder recovered by RSU from quantized bits, where function $\mathcal{D}:\{\pm1\}^Q \rightarrow \mathbb{C}^{N\times1}$ is the method to recover approximate precoding vectors from quantized bits. Three widely-used sensors are considered in this work: Global Positioning System (GPS), RGB cameras, and LiDAR. $\mathbb{M}_k$ denotes the set of informative representations obtained from the $k$-th vehicle's available sensory data, i.e. $\mathbb{M}_k = \{ \mathbf{x}_k^{\rm GPS}, \mathbf{x}_k^{\rm RGB},\mathbf{x}_k^{\rm LiDAR}\}$ if the $k$-th vehicle is equipped with three types of sensors.
% For simplicity,  all subscripts $k$ denote the $k$-th vehicle and will not be restated in the following content. 
The optimization objective of this work is the parameter sets of all local models, both during offline training and online updating processes. Fig.~\ref{wholesystem} depicts the schematic of the system model.

% \vspace{-0.6em}
\section{VFL-Based H-MVMM Precoding Scheme}
\label{HMVMM}
In this section, we first introduce the proposed preprocessing modules for raw multi-modal sensory data. Then, we detail the offline training procedure of H-MVMM.
% \vspace{-1.0em}
\subsection{Multi-Modal Sensing Data Preprocessing}
Due to the lack of explicit correlation between raw sensory data and RSU precoding, directly feeding raw data into NNs leads to learning difficulties. Hence, we tailor three data preprocessing modules for different sensing modalities to convert them into formats that exhibit explicit correlations with various key characteristics relevant to precoding, which will be introduced in the following subsections. 
\begin{figure*}[!t]
	% \centering
	\includegraphics[width=1\linewidth]{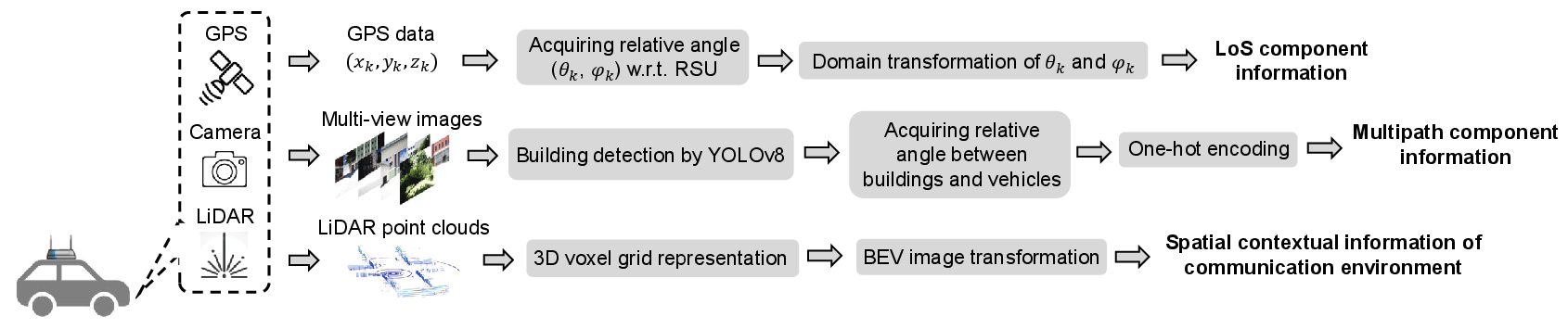}
	\caption{Processing flow of raw multi-modal sensory data.}
    \label{sensingdata}
        % \vspace{-0.5em}
\end{figure*}
\subsubsection{GPS Preprocessing}
Intuitively, GPS indicates the rough location of receivers, providing the LoS component information for RSU. We devise the following preprocessing scheme tailored for GPS data, as shown in Fig.~\ref{sensingdata}.

Without loss of generality, we assume that GPS data is first processed into Cartesian coordinate form $\mathbf{p}_k^t = (x^t_k,y^t_k,z^t_k)^{\rm T}$, with Gaussian noise added to simulate localization errors.   To better align with real-world scenarios, we also account for the issue of GPS signals disappearing due to obstruction by tall buildings.\footnote{Under such condition, the vehicle continuously generates simulated GPS data based on the most recent available GPS data until the GPS signal is restored. It is assumed that the vehicle can continuously obtain its instantaneous velocity to calculate displacement.}

Based on GPS data along with easily accessible RSU's location information $\mathbf{p}_{\rm RSU} = (x_{\rm RSU},y_{\rm RSU},z_{\rm RSU})^{\rm T}$, the vehicle computes the elevation and azimuth angles between itself and RSU as:
\begin{equation}
\label{anglecalculation}
\left\{
\begin{aligned}
\theta_k &= \arctan\left( \frac{y_k - y_{\mathrm{RSU}}}{x_k - x_{\mathrm{RSU}}} \right) \\
\phi_k &= \arccos\left( \frac{z_{\mathrm{RSU}} - z_k}{\Vert \mathbf{p}_k - \mathbf{p}_{\mathrm{RSU}} \Vert_2} \right)
\end{aligned}
\right.
.
\end{equation}

Since NNs may not effectively learn the differences between multiple raw position data input in coordinate formats, we utilize a high-frequency function ${J}(\cdot)$~\cite{position encoding} to encode $\theta_k$ and $\phi_k$ into a high-dimensional domain to facilitate NN learning:
\begin{align}
\label{PE}
{J}(p) = [\sin(2^0\pi p),&\cos(2^0\pi p),
\notag
\\& \cdots,\sin(2^{L-1}\pi p),\cos(2^{L-1}\pi p)],
\end{align}
where $L$ represents the encoding dimension and is set to $5$.
Finally, the LoS component information $\mathbf{x}_k^{\rm GPS} \in \mathbb{R}^{20\times1}$ derived from GPS data is obtained as: $\mathbf{x}_k^{\rm GPS} = [{J}(\theta_k),{J}(\phi_k)]^{\rm T} $.

\subsubsection{RGB Image Preprocessing}
RGB images contain rich semantic information, including potential reflectors like buildings contributing to multipath. Yet, without proper preprocessing on the images, NNs may struggle to extract the angular features of the multipath components, which are required for RSU precoding, from raw pixel values. Hence, we devise the following RGB image processing pipeline to convert raw RGB images to coarse multipath angle information as NN input, as shown in Fig.~\ref{sensingdata}. 

In this work, to capture a comprehensive view of the vehicle's surroundings, we assume that cameras are equipped on all sides of the vehicle. Let $\mathcal{I}_k = \{\mathbf{I}_{{\rm f},k},\mathbf{I}_{{\rm b},k}, \mathbf{I}_{{\rm l},k}, \mathbf{I}_{{\rm r},k}\}$ be the image set captured by the $k$-th vehicle, with $\mathbf{I}_{{\rm f},k}$, $\mathbf{I}_{{\rm b},k}$, $\mathbf{I}_{{\rm l},k}$, and $\mathbf{I}_{{\rm r},k}$ being the images captured by cameras mounted on the front, rear, left, and right sides, respectively.

Firstly, we adopt the object detection model YOLOv8~\cite{yolov8} to detect buildings in the images. Let $f_{\rm{YOLOv8}}(\cdot)$ be the YOLOv8 detector that has been trained on our dataset to accurately detect the buildings. Given an image $\mathbf{I}_{c,k} \in \mathbb{R}^{H \times W \times C}$ from image set $\mathcal{I}_k$ as input, where $c \in \{{\rm f}, {\rm b}, {\rm l}, {\rm r}\}$ denotes the camera position, $H$, $W$, and $C$ denote the height, width and the number of color channels, the output bounding boxes and objectness scores of the detected buildings are obtained as:
\begin{gather}
    \label{yolo}
    % \{\mathcal{B}^{\rm f}_{1,k},\mathcal{B}^{\rm f}_{2,k},\cdots,\mathcal{B}^{\rm f}_{M,k}\}=f_{\rm{YOLOv8}}(\mathbf{I}_{{\rm f},k}), 
    {\mathcal{B}_{c,1,k},\mathcal{B}_{c,2,k},\cdots,\mathcal{B}_{c,M_c,k}}=f_{\rm{YOLOv8}}(\mathbf{I}_{c,k}),
\end{gather} 
where $M_c$ denotes the total number of bounding boxes detected by camera $c$ and $\mathcal{B}_{c,i,k} = \{ {u_{c,i,k},v_{c,i,k},w_{c,i,k},h_{c,i,k},s_{c,i,k}}\}$ denotes the parameter set of the $i$-th bounding box from camera $c$. $u_{c,i,k}$ and $v_{c,i,k}$ are the normalized coordinates of the bounding box center, respectively; $w_{c,i,k}$ and $h_{c,i,k}$ are its normalized width and height, respectively; and $s_{c,i,k}$ represents its objectness score.

% Then, we equate the detected buildings to the centers of bounding boxes and calculate their pixel coordinates as: $\mathbf{p}_{i,k}^{\rm I} = (u_{i,k} H, v_{i,k} W, 1)$. Then, we map these pixel coordinates to camera coordinate system to calculate the actual relative angle from the building's center to the vehicle by $\mathbf{p}_{i,k}^{\rm C} = \mathbf{K}_{\rm{in}}^{-1} \mathbf{p}_{i,k}^{\rm I}, \enspace i=1,\cdots,M$, where $\mathbf{K}_{\rm{in}}$ represents the intrinsic matrix of camera. Then, $\mathbf{p}_{i,k}^{\rm C}$ is normalized by $\widetilde{\mathbf{p}}_{i,k}^{\rm C} = \frac{\mathbf{p}_{i,k}^{\rm C}}{\mathbf{p}_{i,k}^{\rm C}(3)}$ and the azimuth angle between the building's center and vehicle can be obtained by: $\omega_{i,k} = \arctan(\widetilde{\mathbf{p}}_{i,k}^{\rm C}(1))$. 
Then, we equate the detected buildings to the centers of bounding boxes and calculate their pixel coordinates as: $\mathbf{p}_{c,i,k}^{\rm I} = (u_{c,i,k} H, v_{c,i,k} W, 1)$. Then, we map these pixel coordinates to camera coordinate system to calculate the actual relative angle from the building's center to the vehicle by $\mathbf{p}_{c,i,k}^{\rm C} = \mathbf{K}_{\rm{in}}^{-1} \mathbf{p}_{c,i,k}^{\rm I}, \enspace i=1,\cdots,M_c$, where $\mathbf{K}_{\rm{in}}$ represents the intrinsic matrix of camera. Then, $\mathbf{p}_{c,i,k}^{\rm C}$ is normalized by $\widetilde{\mathbf{p}}_{c,i,k}^{\rm C} = \frac{\mathbf{p}_{c,i,k}^{\rm C}}{\mathbf{p}_{c,i,k}^{\rm C}(3)}$ and the azimuth angle between the building's center and vehicle can be obtained by: $\omega_{c,i,k} = \arctan(\widetilde{\mathbf{p}}_{c,i,k}^{\rm C}(1))$. 

\begin{figure}[!t]
	\centering
	\includegraphics[width=1\linewidth]{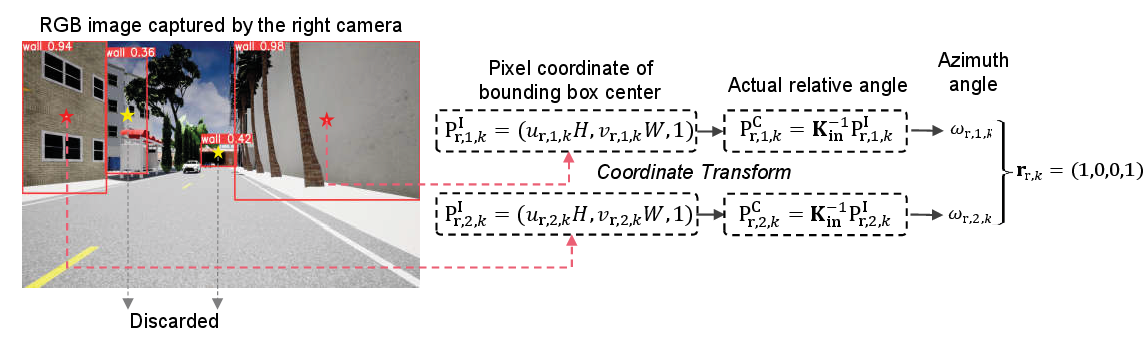}
	\caption{Acquisition process of the indicator vector.
		\label{rgb_process}}
	        % \vspace{-1em}
\end{figure}
Following that, we propose to process the azimuth angles of all detected buildings' centers within an image into an informative one-hot vector indicating the angles of all potential multipath relative to vehicle. Note that we only consider detection results with an objectness score larger than $0.5$. Let $\mathbf{r}_{c,k}$ be the informative vector obtained from camera $c$, where $c \in \{{\rm r}, {\rm l}, {\rm b}, {\rm f}\}$ corresponds to right, left, rear, and front cameras, respectively. We first define a logical variable $q_{c,j,i}$ for each camera $c$:
\begin{equation}
    \label{indicator}
    q_{c,j,i}=\left\{
    \begin{array}{cl}
        0,  & \lfloor \frac{\omega_{c,i,k}-\omega_{\text{min}}}{\Delta \omega} \rfloor = j  ~ \&\&  ~s_{c,i,k}>0.5,   \\
        1,  &  \text{otherwise}
    \end{array} \right.,
\end{equation}
where $\omega_{\text{min}}$ is the minimum azimuth angle within the camera's horizontal field of view, and $\Delta \omega$ is the angle interval, and $j = 1,2,3,4$ indexes the angle bins. Then, $\mathbf{r}_{c,k}$ is calculated by: $\mathbf{r}_{c,k}(j) = 1- \prod \limits_{i=1}^{M_c} q_{c,j,i},~j=1,2,3,4$. This formulation ensures that if any building falls into angle bin $j$, the corresponding element $\mathbf{r}_{c,k}(j)$ is set to 1. Subsequently, the indicator vector $\mathbf{r}_k$ of all potential multipath angles relative to vehicle is derived by concatenating vectors from all cameras: $\mathbf{r}_k = [\mathbf{r}_{{\rm r},k},\mathbf{r}_{{\rm l},k},\mathbf{r}_{{\rm b},k},\mathbf{r}_{{\rm f},k}] \in \mathbb{R}^{1\times16}$. For clarity, the above processing flow is shown in Fig. \ref{rgb_process}. 

Note that $\mathbf{r}_k$ only denotes the angle of the buildings relative to the vehicle itself. Hence, the orientation $\beta_{k}$ of vehicle is another crucial element that allows RSU to learn the absolute angle information about the multipath, which can be easily measured by vehicular sensors like inertial measurement units. Similar to the processing of GPS data, we also encode $\beta_{k}$ using the high-frequency function ${J}(\cdot)$ with the encoding dimension set to $10$. Finally, the multipath component information offered by RGB images is formed as: $\mathbf{x}_k^{\rm RGB} = [ \mathbf{r}_k,{J}(\beta_{k}) ]^{\rm T}  \in \mathbb{R}^{36 \times 1}$.

\subsubsection{LiDAR Preprocessing}
\begin{figure}[!t]
	\centering
	\includegraphics[width=1\linewidth]{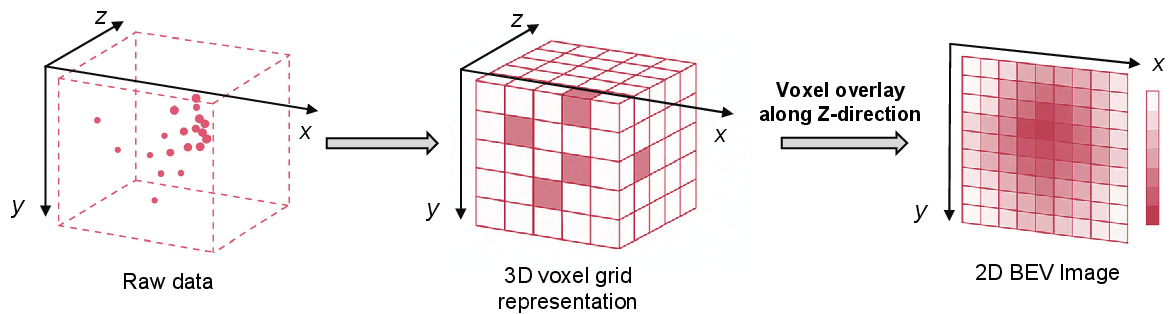}
	\caption{Lightweight processing of LiDAR point cloud.
		\label{lidar_process}}
\end{figure}

LiDAR point clouds collected by multiple vehicles collaboratively depict the spatial structural information of the communication environment, potentially helping RSU model the complex relationships among users through NNs. However, inputting raw LiDAR data into NNs results in extremely high complexity, leading to prolonged NN inference time. To mitigate this, we perform the following lightweight processing, as depicted in Fig.~\ref{lidar_process}.

Firstly, we represent the entire point cloud through a 3D voxel grid $\mathbf{Z}_k \in \mathbb{R}^{L_{\rm x}\times L_{\rm y}\times L_{\rm z}}$. If a voxel in $\mathbf{Z}_k$ contains at least one point of the entire point cloud, it is occupied and marked as $1$. Otherwise, the voxel's value is set to $0$. Next, a bird's eye view  (BEV) image $\mathbf{X}^{\rm LiDAR}_k \in \mathbb{R}^{L_{\rm x}\times L_{\rm y}}$ is formed based on $\mathbf{Z}_k$ by overlaying the values of all voxels in the z-direction and treating the result as the pixel value. Finally, $\mathbf{X}^{\rm LiDAR}_k$ contains sufficient structural information of the vehicle's surrounding environment and is regarded as the input of the LiDAR branch.

It is important to note that the preprocessing described above is simpler than that designed for GPS data and RGB images. This is due to the significant correlation that LiDAR raw data already has with the precoding task, allowing NNs to effectively extract relevant features without requiring complex processing. In contrast, raw GPS data and RGB images lack an explicit relationship with wireless channel. Therefore, intricate preprocessing is necessary to convert them into informative representations that are directly related to the precoding task. 

%accumulating all occupied voxels of $\mathbf{Z}$ in z-direction and
% \vspace{-1em}
\subsection{Offline Training Procedure for H-MVMM}
The offline training procedure of the proposed H-MVMM scheme is illustrated in Algorithm 1. In the $j$-th iteration, each vehicle calculates its local output {\color{black}$\hat{\mathbf{v}}_k = \eta^{-1}\Big(\mathcal{G}_k\big(\eta (\mathbf{y}_k), \mathbb{M}_k\big)\Big)$} and transmits it to RSU, corresponding to line 4 of Algorithm 1. The local model $\mathcal{G}_{k}(\cdot)$ is composed of several uni-modal feature extraction NNs and a multi-modal feature integration NN. Let $\mathcal{G}^{\rm G}(\cdot)$, $\mathcal{G}^{\rm R}(\cdot)$, $\mathcal{G}^{\rm L}(\cdot)$, and $\mathcal{G}^{\rm P}(\cdot)$ be the feature extraction NN designed for GPS, RGB images, LiDAR, and pilots. Then, the multi-modal features are extracted as:
\begin{equation}
    \renewcommand{\arraystretch}{1.2} % 调整行间距，1.5为倍率
    \begin{array}{l l}
        \mathbf{m}_k^{\rm G} = \mathcal{G}^{\rm G}(\mathbf{x}^{\rm GPS}_k), & \mathcal{G}^{\rm G}: \mathbb{R}^{20 \times 1} \mapsto \ \mathbb{R}^{L_{\rm G}\times 1},\\
        \mathbf{m}_k^{\rm R} = \mathcal{G}^{\rm R}(\mathbf{x}^{\rm RGB}_k), & \mathcal{G}^{\rm R}: \mathbb{R}^{36 \times 1} \mapsto \ \mathbb{R}^{L_{\rm R}\times 1}, \\
        \mathbf{m}_k^{\rm L} = \mathcal{G}^{\rm L}(\mathbf{X}^{\rm LiDAR}_k), & \mathcal{G}^{\rm L}: \mathbb{R}^{L_{\rm x} \times L_{\rm y}} \mapsto \ \mathbb{R}^{L_{\rm L}\times 1}, \\
        \mathbf{m}_k^{\rm P} =  \mathcal{G}^{\rm P}(\eta \big(\mathbf{y}_k)^{\rm T}\big), & \mathcal{G}^{\rm P}: \mathbb{R}^{2L_{\rm P}\times 1} \mapsto \ \mathbb{R}^{L_{\rm S}\times 1}. 
    \end{array}
\end{equation}
where $\mathbf{m}_k^{\rm G}$, $\mathbf{m}_k^{\rm R}$, $\mathbf{m}_k^{\rm L}$, and $\mathbf{m}_k^{\rm P}$ denote the uni-modal features obtained from GPS, RGB image, LiDAR, and received pilots, respectively. Assume that the $k$-th vehicle is equipped with three sensors. Then, the vehicle concatenates the uni-modal features as $\mathbf{m}_k^{\rm I} = \mathbf{m}_k^{\rm G} \oplus \mathbf{m}_k^{\rm R} \oplus \mathbf{m}_k^{\rm L} \oplus \mathbf{m}_k^{\rm P}$. Finally, a multi-modal feature integration NN $\mathcal{G}^{\rm I}(\cdot)$ is used to output the precoding vector for the $k$-th user leveraging $\mathbf{m}_k^{\rm I}$:
\begin{equation}
\label{v}
    \hat{\mathbf{v}}_k = \eta^{-1}\big(\mathcal{G}^{\rm I}(\mathbf{m}_k^{\rm I})\big), \enspace \enspace \mathcal{G}^{\rm I}: \mathbb{R}^{(L_{\rm G}+L_{\rm R}+L_{\rm L}+L_{\rm S})\times 1} \mapsto  \mathbb{R}^{2N\times 1}.
\end{equation}

After RSU receives the feedback bits from the vehicles, as described in lines 6 and 7, it evaluates the loss using $\{\mathbf{v}_k\}_{k=1}^K$ as follows: 

\begin{equation}
\label{loss}
\mathcal{L} = -\sum_{k=1}^K (R_k + \lambda^{(R_{\rm T}-R_k)}).
\end{equation}
{\color{black}Notably, the heterogeneity of the local datasets poses challenges for training H-MVMM. Specifically, consider a scenario where one user only has received signals and GPS data, while other users have access to the LiDAR modality. The differences in sensor configurations lead to variations in the sizes of their local NNs. During the early stage of training, the smaller local NNs are scarcely updated. Since the smaller NNs only begin to update after the larger local NNs have nearly converged, the overall training process is slow. To accelerate convergence and address this challenge caused by dataset heterogeneity, we propose a customized loss function, as shown in Eq.~\eqref{loss}.} 
In this loss function, alongside the reciprocal of the sum rate commonly used in existing works, a specially designed regularization term is included. This term reflects the gap between each user's achievable rate $R_k$ and the threshold $R_{\rm T}$. 
% We propose to represent this gap using an exponential function with base $\lambda$. 
We set $R_{\rm T}$ to a small value and $\lambda$ to a large value. This ensures that the regularization term produces a larger derivative during the early stage of training, thereby accelerating the convergence of the small local models. As $R_k$ increases, the derivative of the regularization term approaches zero, ensuring it does not interfere with the optimization objective of maximizing the system sum rate.
% During the initial stage of network training, a large gap $(R_{\rm T}-R_k)$ results in a large derivative, effectively accelerating the convergence speed of smaller local models. 

Subsequently, as described in line 7, RSU calculates the gradients $\big\{\frac{ \partial \mathcal{\mathcal{L}} }{  \partial\mathbf{v}_k} \big\}_{k=1}^K$ for each vehicle and transmits them back to the respective vehicles. Subsequently, each vehicle computes the gradient of its local model $\mathbf{\Theta}_k^j$ and updates it, as shown in lines 9 and 10, where $\eta^{\rm off}_j$ represents the learning rate in the $j$-th iteration. Since the decoding strategy $\mathcal{D}(\cdot)$ is not the ideal inverse mapping of the quantization strategy $\mathcal{F}(\cdot)$,
$\nabla_{\mathbf{\Theta}_k^{j}}\mathcal{L}$ inevitably introduces errors, resulting in an inherent performance loss in the VFL-based scheme. The above process iterates until the entire NN converges. 
% ------------------------------------------------------------
% Algorithm 1 : Offline federated training of H-MVMM
% ------------------------------------------------------------

\begin{algorithm}[t]
\color{black}
\caption{Offline Federated Training of H--MVMM}
\label{alg:offline}
\renewcommand{\algorithmicrequire}{\textbf{Input:}}
\renewcommand{\algorithmicensure}{\textbf{Output:}}
\begin{algorithmic}[1]
\Require Local datasets $\!\bigl\{\mathbf y_k,\mathbb M_k\bigr\}_{k=1}^{K}$;\
        training epochs $N_{\mathrm{off}}$
\Ensure  Precoding matrix $\mathbf V$;\
        trainable local parameters $\{\boldsymbol\Theta_k^{N_{\mathrm{off}}}\}_{k=1}^{K}$
\Statex \textit{//~~Initialization}
\For{$k=1,\dots,K$} \Comment{run on each vehicle}
    \State $\boldsymbol\Theta_k^{0}\leftarrow$ randomly initialize
\EndFor
% ------------------------------------------------------------ \Comment{training epochs}
\For{$j=1,\dots,N_{\mathrm{off}}$}             
    \Statex \textit{//~~1) Local forward \& quantization upload}
    \ForAll{vehicle $k$ \textbf{in parallel}}
        \State $\hat{\mathbf v}_k\gets  \eta^{-1}\Bigl(\mathcal G_k\!\bigl(\eta(\mathbf y_k),\mathbb M_k;\boldsymbol\Theta_k^{j-1}\bigr)\Bigr)$
        \State $\mathbf b_k\gets\mathcal F(\hat{\mathbf v}_k)$ \Comment{quantization}
        \State \textbf{send} $\mathbf b_k$ to RSU
    \EndFor
    % --------------------------------------------------------
    \Statex \textit{//~~2) RSU-side aggregation}
    \State $\mathbf V\gets\bigl[\mathcal D(\mathbf b_1),\dots,\mathcal D(\mathbf b_K)\bigr]$
    \State $\mathcal L\gets\displaystyle
           -\sum_{k=1}^{K}(R_k
           +\lambda^{(R_T-R_k)})$ \hfill(Eq.\,11)
    \State Compute $g_k\gets\partial\mathcal L/\partial\mathbf v_k,\;k=1,\dots,K$
    \State \textbf{unicast} $g_k$ to vehicle $k$
    % --------------------------------------------------------
    \Statex \textit{//~~3) Local backward \& parameter update}
    \ForAll{vehicle $k$ \textbf{in parallel}}
        \State $\displaystyle
               \nabla_{\boldsymbol\Theta_k^{j-1}}\mathcal L
               =g_k\frac{\partial\hat{\mathbf v}_k}{\partial\boldsymbol\Theta_k^{j-1}} $
        \State $\boldsymbol\Theta_k^{j}\gets
               \boldsymbol\Theta_k^{j-1}-\eta^{\mathrm{off}}_k\,
               \nabla_{\boldsymbol\Theta_k^{j-1}}\mathcal L$
    \EndFor
\EndFor
\end{algorithmic}
\end{algorithm}

\section{Label-Free Online Model Updating}
\label{SSLAK}
Section \ref{HMVMM} has introduced the offline training details of the H-MVMM scheme. In this section, we present the details of the label-free online model updating strategy designed for H-MVMM. The main concept involves simulating pseudo DL CSI labels using the PCSI-Simulator. We begin by detailing the step-by-step workflow of the PCSI-Simulator, followed by an explanation of the label-free training methodology.
\begin{figure*}[!t]
	\centering
	\includegraphics[width=1\linewidth]{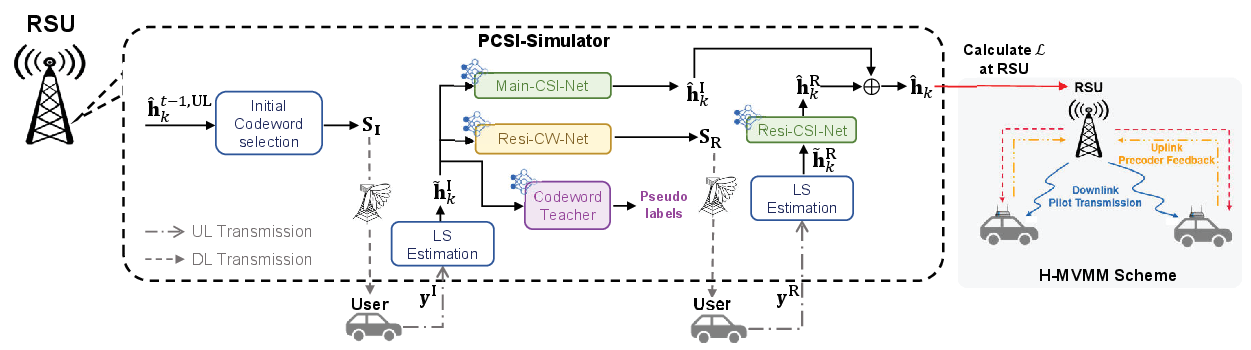}
	\caption{The architecture of PCSI-Simulator.
	\label{onlineupdate}}
\end{figure*}

\subsection{Workflow of PCSI-Simulator}
The workflow consists of three core steps: initial training codeword selection, residual training codeword selection, and pseudo DL CSI label synthesis, as shown in Fig.~\ref{onlineupdate}.

\textbf{{[Step 1] Initial Training Codeword Selection via Partial Reciprocity:}}
The FDD UL and DL channels exhibit partial reciprocity through frequency- and time-invariant multipath parameters, such as angles of arrival (AoAs), angles of departure (AoDs), and path gains \cite{partial-reci1}. Leveraging such partial reciprocity, we propose to directly regard the codewords whose steering angles align with the detected UL AoAs as the initial pilot sequences. Specifically, the approximate AoAs of UL CSI estimates from the previous time slot $\hat{\mathbf{h}}_k^{t-1, \text{UL}} \in \mathbb{C}^{N \times 1}$ are identified using the discrete Fourier transform (DFT) codebook. This can be expressed as: $\hat{\mathbf{e}}_k^{t-1, \text{UL}} = (\hat{\mathbf{h}}_k^{t-1, \text{UL}})^{\rm H}\mathbf{F}$,
where $\hat{\mathbf{e}}_k^{t-1, \text{UL}} \in \mathbb{C}^{1\times N}$ represents the beamspace UL CSI estimates and $\mathbf{F}=[\mathbf{a}(N,0),\mathbf{a}(N,\frac{1}{N}),\cdots,\mathbf{a}(N,\frac{N-1}{N})] \in \mathbb{C}^{N\times N}$ represents the DFT codebook. Here, $\mathbf{a}(N,\phi)\triangleq\frac{1}{N}[1,e^{-j2\pi\phi},\cdots,e^{-j2\pi\phi(N-1)}]^{\rm H}$ denotes the steering vector.

Based on the beamspace UL CSI $\hat{\mathbf{e}}_k^{t-1, \text{UL}}$, we identify the indices of codewords corresponding to the top-$M_1$ largest magnitudes, forming the codeword index set $\mathcal{I}_k^{\rm I}$:
\begin{equation}
\begin{aligned}
    \mathcal{I}_k^{\rm I} = \Big\{ n \in \{1,\cdots,N\} \Big|  \big|\hat{\mathbf{e}}_k^{t-1, \text{UL}}[n] \big|  
    \geq {\text{top-}M_1}\big(\big|\hat{\mathbf{e}}_k^{t-1, \text{UL}}\big|\big)\Big\}. 
\end{aligned}
\end{equation}
Then, the codeword index sets selected for all users are aggregated into $\mathcal{I}_{\rm I} =  \bigcup_{i=1}^{K}\mathcal{I}_i^{\rm I}$. Finally, the DL initial training codewords employed by the RSU $\mathbf{S}_{\rm I}\in\mathbb{C}^{N\times KM_1}$ are constructed by: $\mathbf{S}_{\rm I} = \mathbf{F}[:, \mathcal{I}_{\rm I}]$.\footnote{Here, we assume that the codeword index sets corresponding to different users have no overlapping elements for simplicity. In practice, if there are overlapping elements between different sets, i.e., $\mathcal{I}^{\rm I}_i \cap \mathcal{I}^{\rm I}_j \neq \varnothing, \forall i,j \in \{1,2,\cdots,K\}$, RSU only needs to broadcast the overlapping codewords once. Consequently, the number of codewords transmitted in the first phase will be less than $KM_1$. The construction of $\mathbf{S}_2$ follows the same principle.}

After RSU broadcasts $\mathbf{S}_{\rm I}$, the signals received by the $k$-th user, denoted as $\mathbf{y}_k^{\rm I} \in \mathbb{C}^{1 \times KM_1}$, are expressed as:
\begin{equation}
\label{y}
    \mathbf{y}_k^{\rm I} = \mathbf{h}_k^{\rm H}\mathbf{S}_{\rm I} + \mathbf{n}^{\rm I}_k,
\end{equation}
where $\mathbf{n}^{\rm I}_k \sim \mathcal{CN}(\mathbf{0},\sigma^2\mathbf{I}_{KM_1})\in \mathbb{C}^{1 \times KM_1}$ represents the additive white Gaussian noise. After receiving the signals, the users send them back to RSU for the first round of coarse CSI estimation via the widely used LS algorithm. This can be mathematically expressed as: $ \tilde{\mathbf{h}}_k^{\rm I} = \mathbf{y}^{\rm I}_k(\mathbf{S}_{\rm I}^{\rm H}\mathbf{S}_{\rm I})^{-1}\mathbf{S}_{\rm I}^{\rm H}$.

\textbf{{[Step 2] Residual Training Codeword Selection for CSI Refinement:}} The second round of codeword selection serves two main purposes: first, to identify optimal codewords if those selected in the first stage prove ineffective; and second, to transmit additional suboptimal codewords that accommodate the residual CSI when the initial selections are effective. To achieve these objectives, we propose the construction of the Resi-CW-Net.

Let $\mathcal{G}_{\rm CW}(\cdot)$ be the Resi-CW-Net and $\Theta_{\rm CW}$ be its weight, which takes the coarse LS estimates $\tilde{\mathbf{h}}_k^{\rm I}$ as input. It outputs the probabilities of each codeword in $\mathbf{F}$ being selected by setting the activation function of the output layer to Sigmoid function $f_{\text{Sigmoid}}(x)=\frac{1}{1+e^{-x}}$, i.e., $\hat{\mathbf{p}}_k = \mathcal{G}_{\rm CW}(\eta(\tilde{\mathbf{h}}_k^{\rm I})) \in \mathbb{R}^{N \times 1}$. To avoid duplication between the residual codewords and the initial codewords that are already transmitted, we first set the elements in $\hat{\mathbf{p}}_k$ corresponding to the indices of previously transmitted codewords to zero, i.e., $\hat{\mathbf{p}}_k[i]=0, i\in \mathcal{I}_1$. Afterwards, the set of codeword indices $\mathcal{I}^{\rm R}_k$ corresponding to the largest $M_2$ elements in $\hat{\mathbf{p}}_k$ can be defined as: 
\begin{equation}
    \mathcal{I}_k^{\rm R} = \Big\{ n \in \{1,2,\cdots,N\} \bigg|   \hat{\mathbf{p}}_k[n]   \geq {\text{top-}M_2}\left(\hat{\mathbf{p}}_k\right) \Big\} .
\end{equation}
Subsequently, the codeword index sets of all users can be expressed as: $\mathcal{I}_{\rm R} =  \bigcup_{i=1}^{K}\mathcal{I}_i^{\rm R}$. Finally, the residual training codewords $\mathbf{S}_{\rm R} \in  \mathbb{C}^{N \times KM_2}$ are constructed by: $\mathbf{S}_{\rm R} = \mathbf{F}[:, \mathcal{I}_{\rm R}]$.

\textbf{{[Step 3] Pseudo DL CSI Label Synthesis:}} 
After obtaining $\mathbf{S}_{\rm R}$, RSU performs another round of DL training. The signals received by the $k$-th user $\mathbf{y}_k^{\rm R} \in \mathbb{C}^{1\times KM_2}$ can be expressed as:
\begin{equation}
    \mathbf{y}_k^{\rm R} = \mathbf{h}_k^{\rm H}\mathbf{S}_{\rm R} + \mathbf{n}^{\rm R}_k,
\end{equation}
where $\mathbf{n}^{\rm R}_k \sim \mathcal{CN}(\mathbf{0},\sigma^2\mathbf{I}_{KM_2})\in \mathbb{C}^{1 \times KM_2} $ represents the additive white Gaussian noise. Upon receiving the signals, users feed them back to RSU for coarse estimation of the residual CSI, i.e., $\tilde{\mathbf{h}}_k^{\rm R}=\mathbf{y}^{\rm R}_k(\mathbf{S}_{\rm R}^{\rm H}\mathbf{S}_{\rm R})^{-1}\mathbf{S}_{\rm R}^{\rm H}$.

Then, RSU processes the coarse main CSI estimates $\tilde{\mathbf{h}}_k^{\rm I}$ and residual CSI estimates $\tilde{\mathbf{h}}_k^{\rm R}$ through the Main-CSI-Net and Resi-CSI-Net for CSI refinement. Let $\mathcal{G}_{\rm MC}(\cdot)$ and $\mathcal{G}_{\rm RC}(\cdot)$ be the Main-CSI-Net and Resi-CSI-Net, respectively, with $\Theta_{\rm MC}$ and $\Theta_{\rm RC}$ as their respective weights. Then, the refined CSI can be expressed as: $\eta(\hat{\mathbf{h}}^{\rm I}_k)=\mathcal{G}_{\rm MC}(\eta(\tilde{\mathbf{h}}_k^{\rm I}))$ and $\eta(\hat{\mathbf{h}}^{\rm R}_k)=\mathcal{G}_{\rm RC}(\eta(\tilde{\mathbf{h}}_k^{\rm R}))$. Finally, PCSI-Simulator generates the DL CSI labels by adding $\hat{\mathbf{h}}^{\rm I}_k$ and $\hat{\mathbf{h}}^{\rm R}_k$:
\begin{equation}
    \hat{\mathbf{h}}_k = \hat{\mathbf{h}}^{\rm I}_k + \hat{\mathbf{h}}^{\rm R}_k.
\end{equation}
These labels are subsequently utilized by RSU to calculate the loss using Eq.~\eqref{loss} during the online model update period.

\subsection{Training Methodology of PCSI-Simulator}
In this section, we introduce the label-free training methodology proposed for the PCSI-Simulator. We first present the total loss function used to train the PCSI-Simulator. The total loss $\mathcal{L}_{\rm o}$  is the combination of the MSELoss term $\mathcal{L}_{\rm c}$ for evaluating the accuracy of codewords selected by Resi-CW-Net and the online loss term $\mathcal{L}_{\rm h}$ for measuring the accuracy of generated surrogate CSI labels. Formally, 
\begin{equation}
\label{onlinelosssum}
\mathcal{L}_{\rm o} = \lambda_1\mathcal{L}_{\rm c} + \lambda_2\mathcal{L}_{\rm h}.
\end{equation}
where the weights $\lambda_1$ and $\lambda_2$ are hyperparameters used to balance the two loss terms. For clarity, the procedure of the proposed label-free model online updating strategy is summarized in Algorithm 2. In Algorithm 2, we use $\mathcal{G}_{\rm PCSI}(\cdot)$ to denote the PCSI-Simulator, which comprises multiple NN components, and $\Theta_{\rm PCSI}$ to denote its weights.
Next, we introduce the calculation method for these two loss terms. 

\subsubsection{SSL-Based MSELoss}
We propose to introduce the SSL technique to generate pseudo labels for the supervision of Resi-CW-Net, eliminating the need for a large number of true codeword index labels. The SSL \cite{SSL,SSL2} is a powerful approach that leverages a small amount of labeled data combined with a large amount of unlabeled data to train models. Specifically, we construct a multilayer perceptron (MLP) termed Codeword-Teacher for pseudo label generation. {\color{black} The Codeword-Teacher is designed to output a probability distribution over the optimal codewords, thereby generating pseudo labels for the student network, Resi-CW-Net. Its output format is intentionally aligned with that of the student, enabling direct use of the pseudo labels without additional post-processing. Furthermore, the relative simplicity of the teacher’s task allows it to converge rapidly and maintain robustness when the limited available labels are noisy.  In terms of network design, we keep the Codeword-Teacher’s architecture simple. This is because the task of predicting a probability distribution over candidate codewords is relatively well-defined, and the limited labeled data may contain noise, a complex architecture could easily lead to overfitting.}

\begin{figure}[!t]
	\centering
	\includegraphics[width=1\linewidth]{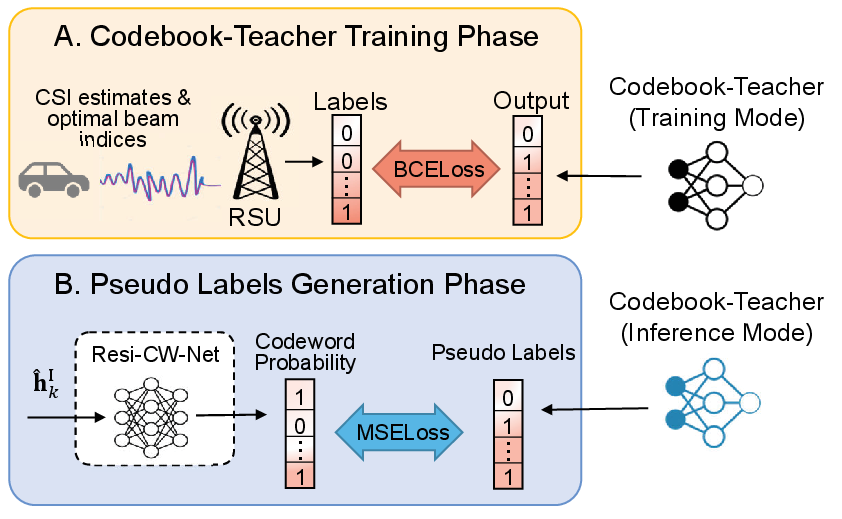}
	\caption{Training and Application of Codeword-Teacher.
	\label{SSLprocess}}
        % \vspace{-1.2em}
\end{figure}

The training method and application strategy of Codeword-Teacher are shown in Fig.~\ref{SSLprocess}. When online updating phase begins, the Codeword-Teacher undergoes supervised training with labeled data until convergence. The process of obtaining labeled data is as follows: we assume that the RSU transmits sufficient downlink pilots so that the user can obtain an accurate estimate of downlink CSI. The user then uses this estimate to compute the optimal beam indices and feeds them back to the RSU as labels. Since the input to the Codebook-Teacher is coarse CSI estimation via LS, users do not need to feedback full CSI estimation but instead transmit the received signals.
 
The training process of the Codeword-Teacher is given as follows.  Let $\mathcal{G}_{\rm T}(\cdot)$ be the Codeword-Teacher and $\Theta_{\rm T}$ be its weight. The Codeword-Teacher takes LS estimates  $\eta(\tilde{\mathbf{h}}_k^{\rm I})$ as input. The activation function of Codeword-Teacher's output layer is set to the Sigmoid function to output the optimal codeword probability, denoted by $\tilde{\mathbf{p}}_k = \mathcal{G}_{\rm T}(\eta(\tilde{\mathbf{h}}_k^{\rm I})) \in \mathbb{R}^{N \times 1}$.
The Binary Cross Entropy (BCE) loss function is utilized to train the Codeword-Teacher:
\begin{equation}
\begin{aligned}
\label{bceloss}
    \mathcal{L}_{\rm BCE}&\left(\{\mathbf{p}_k\}_{k=1}^K,\{\tilde{\mathbf{p}}_k\}_{k=1}^K\right) =
    \\ &\sum_{j=1}^{K}\sum_{i=1}^N\frac{1}{N}\left(  p^i_j\log\tilde{p}^i_j+(1-p^i_j)\log(1-\tilde{p}^i_j) \right),
\end{aligned}
\end{equation}
where $\mathbf{p}_k \in \{0,1\}^N$ denotes the optimal codeword index label in binary form, reported by users through uplink feedback. $p^i_j$ and $\tilde{p}^i_j$ are the $i$-th element of $\mathbf{p}_j$ and $\tilde{\mathbf{p}}_j$, respectively. {\color{black} The BCE loss is adopted because it treats the selection of each candidate codeword as an independent binary decision, which naturally fits the task of identifying the set of optimal beams from the codebook.}

During the training phase of the PCSI-Simulator, the trained Codeword-Teacher operates in inference mode and supplies pseudo labels for the Resi-CW-Net.  We propose to calculate the MSELoss between the output of Resi-CW-Net $\hat{\mathbf{p}}_k$ and pseudo labels $\tilde{\mathbf{p}}_k$ to measure the accuracy of codeword selection results:
\begin{equation}
\label{onlineloss}
\mathcal{L}_{\rm c} = \sum_{k=1}^K \text{MSELoss}(\tilde{\mathbf{p}}_k,\hat{\mathbf{p}}_k) = \sum_{k=1}^K \Vert \tilde{\mathbf{p}}_k - \hat{\mathbf{p}}_k \Vert^2_2.
\end{equation}

\subsubsection{Online Loss}
\label{online loss}
To enable Main-CSI-Net and Resi-CSI-Net to continuously improve their CSI recovery capabilities without relying on true CSI labels, we propose to utilize the following online loss function to supervise them:
\begin{equation}
\label{onlineloss}
\mathcal{L}_{\rm h} = \sum_{k=1}^K \frac{1}{M}\Vert \mathbf{y}^{\rm o}_k - \hat{\mathbf{h}}_k^{\rm H}\mathbf{S} \Vert_1.
\end{equation}
where $M=KM_1+KM_2$ is the total number of DL training codewords, $\mathbf{y}_k^{\rm o}=\mathbf{y}_k^{\rm I}\oplus\mathbf{y}_k^{\rm R} \in \mathbb{C}^{1 \times M}$ represents the entire downlink received signals fed back by users, and $\mathbf{S}=\mathbf{S}_{\rm I} \oplus \mathbf{S}_{\rm R} \in \mathbb{C}^{N\times M}$ represents the overall DL pilot sequences. 

Next, we present an analysis of the continuity and consistency properties pertaining to $\mathcal{L}_{\rm h}$ as follows. Since the $l_1$-norm is a continuous function, the online loss function $\mathcal{L}_{\rm h}$ is continuous with respect to $\hat{\mathbf{h}}_k$. In the following lemma, we will prove that $\mathcal{L}_{\rm h}$ satisfies the consistency requirement, which characterizes the error bound of the generated CSI labels.
% \textit{Lemma 1. [The continuity and convexity of the online loss function]}:
% Lemma 1 indicates that the online loss function $\mathcal{L}_{\rm h}$ is continuous with respect to $\hat{\mathbf{h}}$ and , and is strictly convex with respect to $\hat{\mathbf{h}}$.

% \textit{Proof}: See Appendix \ref{P1}.

\textit{Lemma 1. [Consistency of the online loss function]}:
Given that $\mathbf{S}$ is composed of non-repeating codewords selected from codebook $\mathbf{F}$, $\mathbf{S}$ is column orthogonal such that $\mathbf{S}^{\rm H}\mathbf{S} = {P}I_{M}$. Let $\hat{\mathbf{h}}^{*}$ be the minimizer of the online loss function $\mathcal{L}_{\rm h}$, i.e., 
\begin{equation}
    \hat{\mathbf{h}}^{*}=\mathop{\arg\min}\limits_{\hat{\mathbf{h}}}\frac{1}{M}\Vert \mathbf{y}^{\rm o}-\hat{\mathbf{h}}^{\rm H}\mathbf{S}\Vert_1.
\end{equation}
Then, the error of $\hat{\mathbf{h}}^{*}$ in the $l_2$-norm satisfies the following bound:
\begin{equation}
% \vspace{-0.5em}
    \Vert \hat{\mathbf{h}}^{*} - \mathbf{h} \Vert^2_2 \leq \frac{NM\alpha^2\sigma^2}{P} ,
    % \vspace{-0.5em}
\end{equation}
with probability at least $1-\exp\left( -\frac{N}{2}(\alpha-1)^2 \right)$ for any $\alpha>1$.

\textit{Proof}: See Appendix A in the supplementary material.

Lemma 1 demonstrates that a specific DNN optimized for the online loss function $\mathcal{L}_{\rm h}$ achieves a bounded channel recovery error, with asymptotic convergence to zero as $\text{SNR} \to \infty$.   This convergence occurs because the  parameters $N$ and $M$ are independent of transmit power $P$.

% ------------------------------------------------------------
% Algorithm 2 : Label-free online updating with PCSI-Simulator
% ------------------------------------------------------------
\begin{algorithm}[t]
\color{black}
\caption{Label-Free Online Updating of H--MVMM}
\label{alg:online}
\renewcommand{\algorithmicrequire}{\textbf{Input:}}
\renewcommand{\algorithmicensure}{\textbf{Output:}}
\begin{algorithmic}[1]
\Require Uplink-fed DL signals $\{\mathbf y_k^{\mathrm o}\}_{k=1}^{K}$ of the current time slot,\
        UL CSI estimates from the previous time slot $\hat{\mathbf{h}}^{t-1, \text{UL}}$,\
        pilot sequence $\mathbf S$,\
        local datasets $\{\mathbf y_k,\mathbb M_k\}_{k=1}^{K}$,\
        hyper-parameters $(T_{\mathrm{CT}},T_{\mathrm{PS}},T_{\mathrm{up}})$
\Ensure Updated local parameters $\{\boldsymbol\Theta_k^{T_{\mathrm{up}}}\}_{k=1}^{K}$
% ------------------------------------------------------------
\Statex\textit{//~~Phase A: construct $\mathcal D_1$ (small labeled set)}
\State Obtain $N_{\rm c}$ tuples $\!(\tilde{\mathbf{h}}^{\mathrm{I}},\mathbf p)$ from users' feedback
      and store in $\mathcal D_1$

% ------------------------------------------------------------
\Statex\textit{//~~Phase B: SSL training of Codeword-Teacher}
\For{$j=1,\dots,T_{\mathrm{CT}}$}
    \State $\mathcal L_{\mathrm{BCE}}\!=
           \frac1{|\mathcal D_1|}\!
           \sum_{(\tilde{\mathbf{h}}^{\rm I},\mathbf p)\in\mathcal D_1}\!
           \text{BCELoss}\bigl(\mathcal{G}_{\rm T}(\eta(\tilde{\mathbf{h}}^{\rm I})),\mathbf p\bigr)$
    \State $\boldsymbol\Theta^j_{\mathrm T}\leftarrow
           \boldsymbol\Theta^{j-1}_{\mathrm T}-\eta_{\mathrm T}\nabla_{\Theta^{j-1}_{\mathrm T}}\mathcal L_{\mathrm{BCE}}$
\EndFor

% ------------------------------------------------------------
\Statex\textit{//~~Phase C: construct $\mathcal D_2$ (unlabeled set)}
\State Collect $N_{\rm g}$ tuples $\bigl(\mathbf y_k^{\mathrm o},\mathbf S\bigr)$ and store in $\mathcal D_2$

% ------------------------------------------------------------
\Statex\textit{//~~Phase D: train PCSI-Simulator}
\State $\bigl\{\boldsymbol\Theta_{\rm CW}^{k,0},
             \boldsymbol\Theta_{\rm MC}^{k,0},
             \boldsymbol\Theta_{\rm RC}^{k,0}\bigr\}_{k=1}^{K}
       \leftarrow \text{randomly initialize}$
\For{$m=1,\dots,T_{\mathrm{PS}}$}
    \ForAll{$(\mathbf y^{\mathrm o},\mathbf S)$ in mini-batch}
        \State Generate pseudo DL CSI labels $\hat{\mathbf h}\!=\!\mathcal{G}_{\rm PCSI}(\hat{\mathbf{h}}^{t-1, \text{UL}}, \mathbf y^{\mathrm o};\boldsymbol\Theta_{\mathrm{PCSI}})$
        \State $\mathcal L_c\!=\!
               \text{MSELoss}\bigl(
               \mathcal{G}_{\rm CW}(\eta(\tilde{\mathbf{h}}^{\rm I})),\,
               \mathcal{G}_{\rm T}(\eta(\tilde{\mathbf{h}}^{\rm I}))\bigr)$
        \State $\mathcal L_h\!=\!
               \tfrac1{M}\bigl\lVert \mathbf y^{\mathrm o}-\hat{\mathbf h}^{\mathrm H}\mathbf S\bigr\rVert_1$
        \State $\mathcal L_o=\lambda_1\mathcal L_c+\lambda_2\mathcal L_h$
    \EndFor
    % \State Update $\boldsymbol\Theta_{\mathrm{CW}},\boldsymbol\Theta_{\mathrm{MC}},\boldsymbol\Theta_{\mathrm{RC}}
    \For{ $k=1,2,\cdots,K$ }
          \State $\Theta_{\rm CW} ^{k,m} \leftarrow \Theta_{\rm CW} ^{k,(m-1)} - \eta_{\rm G} \nabla_{\Theta_{\rm CW}^{k,(m-1)}  }\mathcal{L}_{\rm o}$
            \State $\Theta_{\rm MC} ^{k,m} \leftarrow \Theta_{\rm MC} ^{k,(m-1)} - \eta_{\rm G} \nabla_{\Theta_{\rm MC}^{k,(m-1)} }\mathcal{L}_{\rm o}$ 
            \State $\Theta_{\rm RC}^{k,m} \leftarrow \Theta_{\rm RC}^{k,(m-1)} - \eta_{\rm G} \nabla_{\Theta_{\rm RC}^{k,(m-1)} }\mathcal{L}_{\rm o}$ 
    \EndFor
\EndFor
% ------------------------------------------------------------
\Statex\textit{//~~Phase E: federated fine-tuning of H-MVMM}
\For{$k=1,\dots,K$}
    \State Initialize $\boldsymbol\Theta_k^{0}\!\gets\!\boldsymbol\Theta_k^{N_{\mathrm{off}}}$ \Comment{reuse offline weights}
\EndFor
\For{\text{each iteration} $i=1,2,\cdots, T_{\rm up}$}
    \State Execute steps 3 to 11 of Algorithm 1. \Comment{Alg.\,\ref{alg:offline}}
\EndFor
\end{algorithmic}
\end{algorithm}

% \vspace{-1em}
\subsection{Convergence Analysis of PCSI-Simulator}
In this section, we establish a theoretical support for the convergence of the proposed PCSI-Simulator, which essentially learns the inverse mapping $\varphi$ from the coarse LS estimates $ \tilde{\mathbf{h}}_k$ to the true CSI ${\mathbf{h}}_k$, i.e.,
\begin{equation}
        \hat{\mathbf{h}}_k^{*}=\mathop{\arg\min}\limits_{\hat{\mathbf{h}}}\frac{1}{M}\Vert \mathbf{y}_k^{\rm o}-\hat{\mathbf{h}}_k^{\rm H}\mathbf{S}\Vert_1\triangleq \varphi(\tilde{\mathbf{h}}_k).
\end{equation}
Since the PCSI-Simulator optimization process for each user involves updating dedicated network parameters through independent back propagation, we specifically analyze the sub-loss in $\mathcal{L}_{\rm h}$ for an arbitrary user without loss of generality. Therefore, we omit the subscript $k$ in this subsection.We first present the following lemma which bounds the error between the output of the PCSI-Simulator and the inverse mapping based on the universal approximation theorem. 
% We initially derive the representation complexity bound of DNNs that asymptotically approach the minimizer of the online loss functionthe optimal solution $\hat{\mathbf{h}}^{*}$ minimizing the online loss function $\mathcal{L}_{\rm h}$.
 % Furthermore, we use DNN, a universal NN structure, to replace DL-CSI Generator in this section to analyze its convergence. 

\textit{Lemma 2. [Universal Approximation Theorem for Surrogate CSI Label Generation]}:
We assume that the inverse mapping $\varphi$ is continuous in a certain compact set $\mathcal{C}$. Then, the mapping from $ \tilde{{\mathbf{h}}}^{\text{vec}} = \eta(\tilde{\mathbf{h}})\in \mathbb{R}^{2N \times 1}$ to $\hat{\mathbf{h}}^*_{\text{vec}}=\eta(\hat{\mathbf{h}}^*)\in\mathbb{R}^{2N \times 1}$:
\begin{equation}
\label{mapping}
    \hat{\mathbf{h}}^*_{\text{vec}} = \eta(\varphi(\tilde{\mathbf{h}}))\triangleq \varphi^{\prime}(\tilde{{\mathbf{h}}}^{\text{vec}}), 
\end{equation}
is universally approximable by a DNN. Specifically, given any precision threshold $\varepsilon>0$, there exists a DNN $\mathcal{G}_{\text{DNN}}(\cdot)$ with width $W$ and at most $D=2(\lfloor\log_2(2N)\rfloor)$ layers, such that
\begin{equation}
    \sup_{\tilde{\mathbf{h}}^{\rm vec}\in \mathcal{C}} \Vert\mathcal{G}_{\text{DNN}}(\tilde{{\mathbf{h}}}^{\text{vec}})-\varphi^{\prime}(\tilde{{\mathbf{h}}}^{\text{vec}})\Vert_2^2 \leq \varepsilon
\end{equation}
 
\textit{Proof}: See Appendix B in the supplementary material.

Lemma 2 demonstrates that under continuity of the inverse mapping $\varphi$ on a compact set, a DNN with bounded number of layers can universally approximate $\varphi$ with arbitrarily small error. The following Lemma 3 ensures the continuity of the inverse mapping.

\textit{Lemma 3. [Continuity of the Inverse Mapping]}: 
% We assume that $\mathcal{Y}$ and $\mathcal{S}$ are certain compact sets, and $\Gamma = \mathcal{Y} \times \mathcal{S}$. Then, 
Given the online loss function $\mathcal{L}_{\rm h}$ on $\mathcal{C}$, with $\hat{\mathbf{h}} \in \hat{\mathcal{H}}$ and $\tilde{\mathbf{h}} \in {\mathcal{C}}$, the following inverse mapping
\begin{equation}
% \vspace{-0.3em}
\hat{\mathbf{h}}^{*}=\mathop{\arg\min}\limits_{\hat{\mathbf{h}}\in \hat{\mathcal{H}}} \mathcal{L}_{\rm h} = \mathop{\arg\min}\limits_{\hat{\mathbf{h}}\in \hat{\mathcal{H}}} \frac{1}{M}\Vert \mathbf{y}^{\rm o}-\hat{\mathbf{h}}^{\rm H}\mathbf{S}\Vert_1
% , \{\mathbf{y}^{\rm o},\mathbf{S} \} \in \Gamma
% \vspace{-0.3em}
\end{equation}
is a continuous function on $\mathcal{C}$ when being optimized over the compact set $\hat{\mathcal{H}}=\{\hat{\mathbf{h}}:\Vert \hat{\mathbf{h}}\Vert_2^2 \leq C_{\rm h} + \frac{NM\alpha^2\sigma^2}{P}\}$ for some constant $C_{\rm h}$ with probability at least $1-\exp\left( -\frac{N}{2}(\alpha-1)^2 \right)$.

\textit{Proof}: See Appendix C in the supplementary material.

Lemma 3 establishes the continuity of the optimal solution of $\hat{\mathbf{h}}$, i.e., the inverse mapping of the online loss function $\mathcal{L}_{\rm h}$, within the compact set ${\hat{\mathcal{H}}}$. By integrating the properties of $\mathcal{L}_{\rm h}$ given in Section \ref{online loss} with Lemma 2 and 3, we derive the following convergence result for the PCSI-Simulator:
 
\textit{Theorem 1. [Convergence Analysis for PCSI-Simulator]}:
Given any precision threshold $\varepsilon>0$ and online loss function $\mathcal{L}_{\rm h}$, there exists a DNN with width $W$ and at most $D=2(\lfloor\log_2(2N)\rfloor)$ layers, such that the output of the DNN $\mathcal{G}_{\text{DNN}}(\tilde{{\mathbf{h}}}^{\text{vec}})$ satisfies:
\begin{equation}
     % \sup_{(\mathbf{y}^{\rm o},\mathbf{S})\in \Gamma} \| \eta^{-1}(\mathcal{G}_{\text{DNN}}(\tilde{{\mathbf{h}}}^{\text{vec}}))- \mathbf{h}\|^2_2 
     % \leq  \varepsilon + \frac{NM\alpha^2\sigma^2}{P}
     \sup_{\tilde{\mathbf{h}}^{\rm vec}\in \mathcal{C}} \| \eta^{-1}\big(\mathcal{G}_{\text{DNN}}(\tilde{{\mathbf{h}}}^{\text{vec}})\big)- \mathbf{h}\|^2_2 
     \leq  \varepsilon + \frac{NM\alpha^2\sigma^2}{P},
\end{equation}
with a probability of at least $\min\left(1-\epsilon,1-\exp\left( -\frac{N}{2}(\alpha-1)^2\right)\right)$ in the compact set $\mathcal{C} = \{\tilde{\mathbf{h}}^{\rm vec} : \|\tilde{\mathbf{h}}^{\rm vec}\|_2^2 \le C_{\rm h} + \frac{\sigma^2}{P}Q_{\chi^2_{2M}}(1-\epsilon)\}$  for any prescribed parameters $\alpha>1$ and $\epsilon>0$. In this work, the specific implementation of the above mentioned DNN is the proposed PCSI-Simulator.

% $\mathcal{Y}=\{\mathbf{y}^{\rm o}:\|\mathbf{y}^{\rm o}\|_2^2\leq  2PC_h + 2\sigma_n^2Q_{\chi_{2K}^2}(1-\epsilon)\}$ and $\mathcal{S}=\{\mathbf{S}:\|\mathbf{S} \|^2_{\rm F}\leq KP\}$.
\textit{Proof}: See Appendix D in the supplementary material.

% \vspace{-1.6em}
\textit{Remark 1}: Taking the condition where the number of users changes dynamically as an example, we illustrate how the H-MVMM scheme updates the local models through the proposed strategy.

Most existing DL approaches pre-train and store multiple NNs with varying input/output sizes to accommodate changes in user number. However, this strategy proves impractical in the considered scenario, where both the user count and their sensor configurations are variable. For instance, when a new vehicle joins, existing vehicles would need to pre-download numerous models to adapt to the new vehicle's potential local model configurations. Specifically, with $K=6$, accommodating $N_{\rm sensor}$ potential sensor configurations for the new user would require storing up to $N_{\rm sensor} \times K = (C_{3}^{1}+C_{3}^{2}+C_{3}^{3})\times 6 = 42$ pre-trained models in advance.

The aforementioned inflexibility can be effectively mitigated through a dynamic user-adaptive approach, leveraging the online model updating strategy. During the online inference stage, when new vehicles enter, they first download \textit{randomly initialized} models and subsequently train the corresponding NN branches from scratch based on their available sensors. Concurrently, existing vehicles update their local \textit{pre-trained models}. The training procedure adheres to Algorithm 2. This dynamic user-adaptive approach enables the overall NN to converge in a short number of epochs, allowing the system to quickly re-enter the inference stage without relying on true labels or consuming extensive resources to obtain CSI estimates as labels. When existing vehicles leave, the remaining vehicles only need to update their \textit{pre-trained models}.

{\color{black}
\subsection{Complexity Analysis}
In this section, we quantify the computational complexity of the proposed H-MVMM precoding scheme and the label-free online model updating strategy. We first present the complexity analysis of the vehicle-side local NNs. The received pilots branch, the RGB branch, and the GPS branch share similar computational complexity, which is $\mathcal{O}(aL_{\rm f})$, where $a$ denotes the input size and $L_{\rm f}$ represents the output feature size. The LiDAR branch utilizes a convolutional neural network to process LiDAR data, leading to a higher complexity of $O(HWC_{\rm i}C_{\rm o}K_{\rm e}^2)$, where $H$ and $W$ are the input feature map dimensions, $C_{\rm i}$ and $C_{\rm o}$ are the number of input and output channels, and $K_{\rm e}$ is the kernel size. Therefore, the overall complexity of the local NNs is dominated by the LiDAR branch, which is $O(HWC_{\rm i}C_{\rm o}K_{\rm e}^2)$.

The proposed PCSI-Simulator used in online model updating phase consists of four NN models. Among these models, the Main-CSI-Net and the Resi-CSI-Net both employ a 1-layer Transformer encoder, which dominates the complexity. The input projection and output projection layers of these two models each exhibit a complexity of $\mathcal{O}(DL)$. The multi-head self-attention mechanism demonstrates a complexity of $\mathcal{O}(DL^2)$, where $L=N$ is the sequence length and $D$ is the model dimension. Additionally, the feedforward network contributes  $\mathcal{O}(DLF)$, with $F$ being the feedforward dimension. Consequently, given that $F < L$, the dominant operation in the Main-CSI-Net and the Resi-CSI-Net is the self-attention mechanism, with a complexity of $\mathcal{O}(DL^2)$.
The PCSI-Simulator also necessitates two rounds of coarse CSI estimation  via the LS algorithm. In LS estimation, the initial operation involves forming the matrix $\mathbf{S}_{\rm R}^{\rm H}\mathbf{S}_{\rm R}$ with a complexity of $\mathcal{O}(NM^2)$. The subsequent steps include the inversion of $\mathbf{S}_{\rm R}^{\rm H}\mathbf{S}_{\rm R} \in \mathbb{C}^{M \times M}$ matrix with a complexity of $\mathcal{O}(M^3)$, followed by post-multiplication by $\mathbf{S}_{\rm R}^{\rm H}$ with a complexity of $\mathcal{O}(NM^2)$.  Given that $M$ is significantly smaller than $N$ (e.g., $M=9$), the complexity is dominated by the first step, resulting in a per-user complexity of $\mathcal{O}(NM^2)$ for LS channel estimation.

In summary, during the online model updating phase, the total computational complexity is $\mathcal{O}(KHWC_{\rm i}C_{\rm o}K_{\rm e}^2+KDL^2+KNM^2)$.  Since most parameters, such as the model dimension $D$, the number of training codewords $M$, and the kernel size $K_{\rm e}$, are relatively small and fixed in practice, and noting that $L = N$,  the complexity is dominated by the term $KDN^2$, i.e., $\mathcal{O}(KN^2)$. Therefore, the computational overhead is manageable for real-time implementation.
}

\section{Experiment Configurations}
In this section, we present the dataset, outline the implementation details of the proposed scheme, and describe the benchmarks used for evaluation.
\label{ES}
 % \vspace{-1em}
\subsection{Dataset}
\begin{figure}[!t]
	\centering
	\includegraphics[width=1\linewidth]{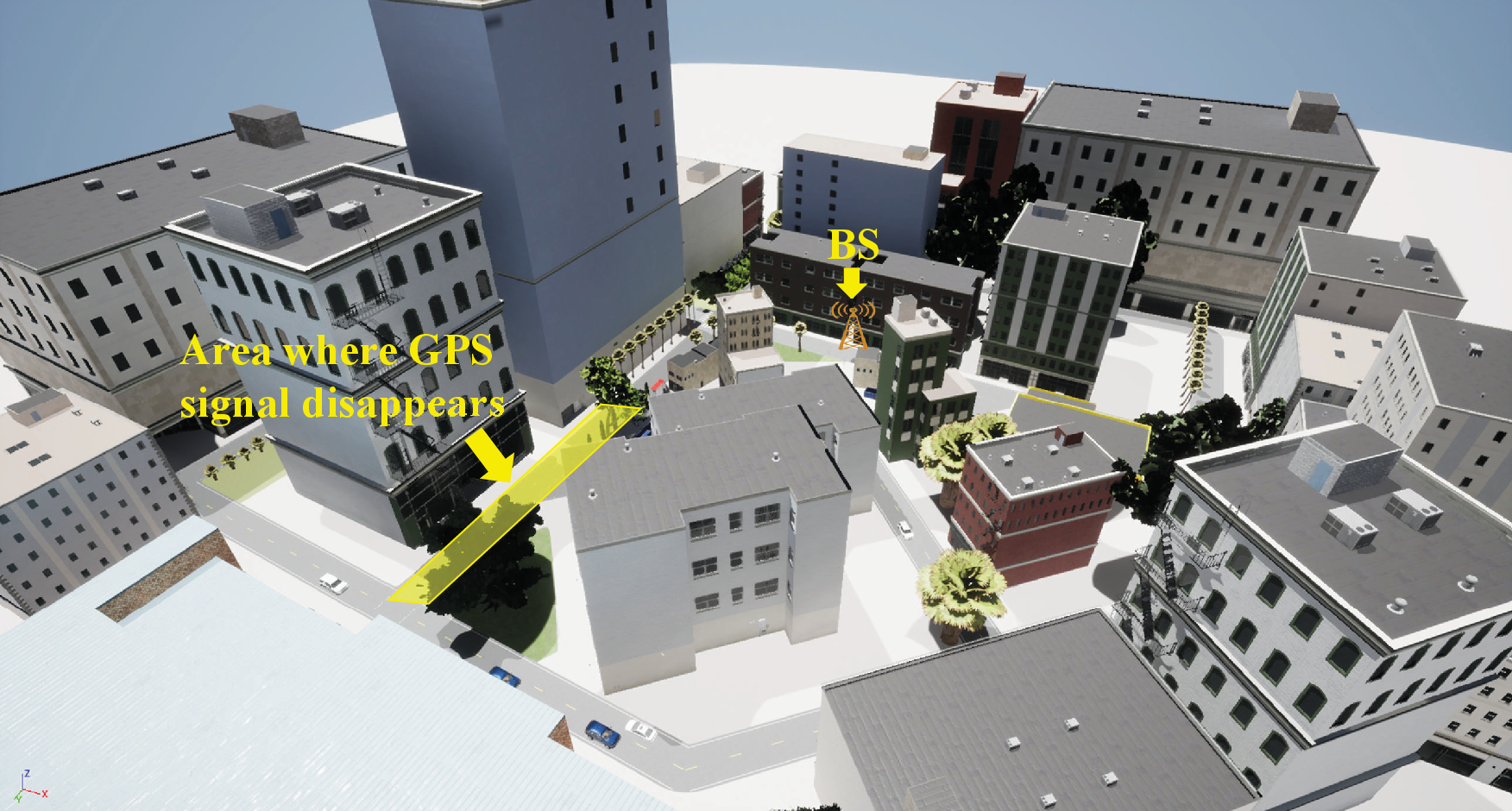}
	\caption{Overview of the city block scenario.
	\label{scenario}}
        % \vspace{-1.2em}
\end{figure}

We evaluate our method using the city block scenario from the SynthSoM dataset~\cite{M3C}, as shown in Fig.~\ref{scenario}. The area where GPS signal disappears and the location of the RSU are also shown. This scenario features packed buildings and many non-line-of-sight (NLoS) communication conditions. The size of the scenario is  $190$m  $\times$ $135$m. The RSU is equipped with $N_{\rm{v}} \times N_{\rm h} = 16 \times 8$ antennas, with the height set to $z_{\rm RSU} = 9$m. The DL and UL carrier frequencies are $4.95$ GHz and $4.85$ GHz, respectively.  The mean and standard deviation of the Gaussian noise added to the x and y coordinates of GPS data are set to $0$ and $5$, respectively. The size of LiDAR BEV is set to $ L_{\rm x} \times L_{\rm y} = 64 \times 64$.
The minimum azimuth angle $\omega_{\text{min}}$ of the camera is $-0.87$rad and the angle interval $\Delta \omega$ is set to $0.435$rad. In the following experiments, all vehicles are assumed to have the pilot modality, with $5$ out of $7$ vehicles equipped with GPS, $3$ vehicles equipped with RGB cameras, and $3$ vehicles equipped with LiDAR. {\color{black}The sensor configuration details for all vehicles are shown in Table \ref{setup}.}

\begin{table}[!t]
	\setlength{\abovecaptionskip}{0.1cm} 
	\renewcommand\arraystretch{0.05} 
	\centering
    \small 
	\caption{Sensor configurations of different vehicles}
	\label{setup}
    {\color{black}
	\begin{tabular}{c|c|c|c|c}
		\toprule[0.35mm]
		\textbf{Vehicle} &\textbf{Pilot}	&\textbf{GPS} &\textbf{RGB Camera}  &\textbf{LiDAR}  \\
		\midrule[0.15mm]
		Vehicle 1	& \checkmark & \checkmark & -- & \checkmark\\ 
		\midrule[0.15mm]
		Vehicle 2 & \checkmark & \checkmark & -- & --\\ 
		\midrule[0.15mm]			
		Vehicle 3	& \checkmark & -- & \checkmark & --\\ 	
		\midrule[0.15mm]			
		Vehicle 4 & \checkmark & -- & -- & \checkmark \\
		\midrule[0.15mm]		
		Vehicle 5  & \checkmark & \checkmark & \checkmark & --\\ 	
		\midrule[0.15mm]		
		Vehicle 6  & \checkmark & \checkmark & -- & \checkmark\\ 
		\midrule[0.15mm]			
		Vehicle 7 & \checkmark & \checkmark & \checkmark & --\\ 	
		\bottomrule[0.35mm]		
	\end{tabular}	
    }
\end{table}

% \vspace{-0.5em}
\subsection{Implementation Details}
\subsubsection{Offline-Trained H-MVMM Scheme}
\begin{figure}[!t]
	\centering
	\includegraphics[width=1\linewidth]{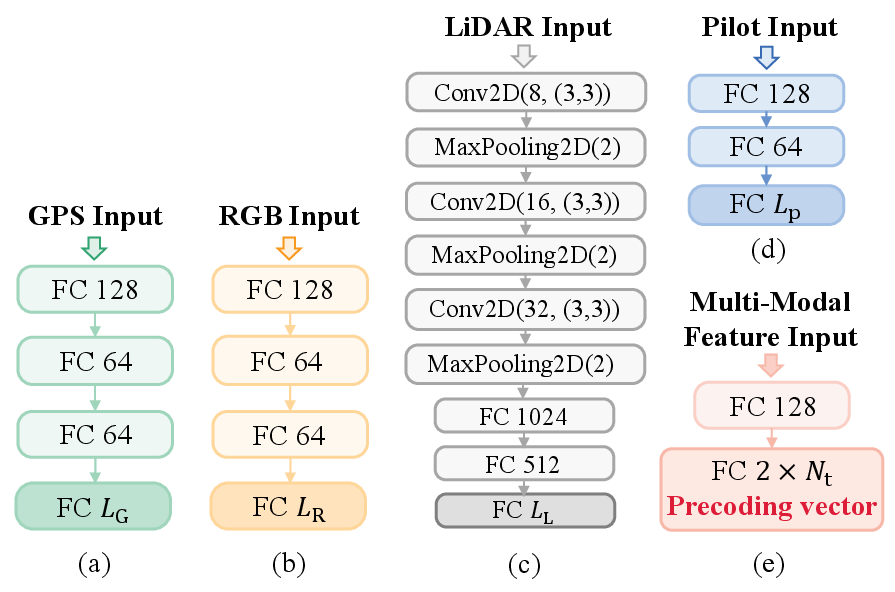}
	\caption{Illustration of NN architectures for: (a) GPS branch; (b) RGB image branch; (c) LiDAR branch; (d) received pilots branch; (e) multi-modal feature integration network.
	\label{NN}}
    % \vspace{-1.5em}
\end{figure}

We use the adaptive moment estimation (ADAM) \cite{Adam} as the optimizer and initialize the learning rate to $\{\eta^{\rm off}_k\}_{k=1}^{K} = 1 \times 10^{-4}$ for the H-MVMM scheme and its uni-modality counterparts. We train different schemes with a batch size of $32$ until the validation loss stabilizes. The architectures of NNs customized for different sensing modality and the multi-modal feature integration network at the \textit{user side} are depicted in Fig.~\ref{NN}. The sizes of multi-modal features are set to $L_{\rm S}=128$, $L_{\rm G}=256$, $L_{\rm R}=256$, and $L_{\rm L}=512$, respectively. The threshold $R_{\rm T}$ in the customized loss function Eq.~\eqref{loss} is set to $0.3$ and the base $\lambda$ is set to $10$. These two parameters have been experimentally proven to be suitable for optimizing the training of local models. The number of quantization bits is set to $B=2$.

\subsubsection{PCSI-Simulator}
The ADAM optimizer is used to update the weights of the PCSI-Simulator and Codeword-Teacher, with the batch size set to $32$ for both. The Main-CSI-Net and Resi-CSI-Net both consist of a $1$-layer Transformer encoder \cite{position encoding} ($16$ heads, $32$-dim embeddings) with $32$-dim feedforward projections and $0.3$ dropout, followed by a $1\times1$ convolutional layer that compresses $32$-channel feature maps to $2$ output dimensions. The Codeword-Teacher and Resi-CW-Net share the same structure, adopting a $3$-layer MLP architecture with batch normalization applied after the first and second layers, where the number of hidden neurons in each layer is $[l_1, l_2, l_3] = [256, 512, 128]$. 

The total number of epochs required for the updating process of H-MVMM $T_{\rm up}$ depends on the SNRs. Specifically, $250$ epochs are required when the SNR is below $15$ dB, $600$ epochs are needed when the SNR is above $18$ dB, and $400$ epochs are used for all other cases. The PCSI-Simulator and Codeword-Teacher undergo training for $T_{\rm PS}=150$ epochs and $T_{\rm CT}=100$ epochs, respectively. Learning rates are initialized at $\eta_{\rm T}=2\times10^{-4}$ for Codeword-Teacher and $\eta_{\rm G}=1\times10^{-3}$ for Main-CSI-Net/Resi-CSI-Net/Resi-CW-Net, with scheduled halving at epochs $[20,40,60,100]$ and $[10,20,40,60,100]$ respectively. During the updating process, the learning rates of local models remain the same as those adopted in the offline training phase. The weights of sub-loss $\lambda_1$ and $\lambda_2$ in Eq. \eqref{onlinelosssum} are both set to~$1$. 

In this work, the Codeword-Teacher is trained on a dataset of size $|\mathcal{D}_1|=KN_{\rm c}$, where $N_{\rm c}=120$ is the number of DL channel estimate samples fed back by each user. The size of dataset $\mathcal{D}_2$ constructed for PCSI-Simulator is configured as $|\mathcal{D}_2|=KN_{\rm g}$ ($N_{\rm g}=1000$), with each data sample comprising $M$ received signals fed back by each user.

% \vspace{-1em}
\subsection{Benchmarks} 
\begin{itemize}
\item[$\bullet$] 
\textbf{ZF w/ GT CSIT}: ZF~\cite{ZF} method with ground truth CSI is used as a benchmark scheme.
\end{itemize}

\begin{itemize}
\item[$\bullet$] 
\textbf{WMMSE w/ GT CSIT}: WMMSE~\cite{WMMSE} method with ground truth CSI is used as a benchmark scheme. It is widely adopted as the {upper bound} of the system sum rate performance as it provides a very decent local optimum.
\end{itemize}

\begin{itemize}
\item[$\bullet$] 
\textbf{RSU-NN}: RSU-NN is a DL-based precoding scheme implemented within the CL framework. RSU-NN uses CSI estimates (CE) as input, with the normalized mean squared error of CE being $-30$dB. Given that the primary goal of pilot transmission is to achieve accurate CSI, the RSU-NN scheme can be considered an approximate upper bound for the performance of the Uni-Pilot scheme.
\end{itemize}
% \vspace{-1em}
\section{Numerical Results}
\label{Expriment}

\begin{table}[!t]
	\setlength{\abovecaptionskip}{0.1cm} 
%	\linespread{1}
	\renewcommand\arraystretch{0.3} 
	\centering
	\caption{Sum rate performances of different offline-trained schemes with respect to $K$ for SNR=$30$dB.}
        \resizebox{1\linewidth}{!}{
	\label{sumrate1}
	\begin{tabular}{@{\hspace{1.2em}}c@{\hspace{1.2em}}|@{\hspace{1.2em}}c@{\hspace{1.2em}}}
		\toprule[0.35mm]
		{\textbf{Scheme}}  &\makecell[c]{\textbf{Sum rate performance} (bps/Hz) \\ (K=$3/4/5/6/7$)}  \\
		% \midrule[0.15mm]
		% \makecell[c]{Uni-Pilot scheme} & \underline{$1.8$}/$84.0$	& \underline{$70.5$}/$74.1$ \\ 
		\midrule[0.15mm]
  		\makecell[c]{Uni-Pilot} & \makecell[c]{$27.3$~/~$36.3$~/~$45.2$~/~$54.4$~/~$65.9$ } \\ 
		\midrule[0.15mm]
		\makecell[c]{Pilot-LiDAR}  & {$33.1$}~/~{$42.2$}~/~{$54.1$}~/~{$66.9$}~/~{$80.2$} \\ 
		\midrule[0.15mm]			
		\makecell[c]{Pilot-RGB} 	& $31.1$~/~$39.6$~/~$50.4$~/~{$60.9$}~/~$72.4$  \\
		\midrule[0.15mm]			
		  \makecell[c]{Pilot-GPS}  & $29.1$~/~$37.3$~/~$45.5$~/~{$56.9$}~/~$68.1$   \\
		\midrule[0.15mm]		
		\makecell[c]{H-MVMM} & $30.5$~/~$40.0$~/~$51.2$~/~{$63.1$}~/~$75.2$  \\ 
             \midrule[0.15mm]		
		\makecell[c]{{\color{black}\textbf{WMMSE w/ GT CSIT }}} & \textbf{{35.2}}~/~\textbf{{45.0}}~/~\textbf{{56.6}}~/~\textbf{{68.7}}~/~\textbf{81.6 } \\ 
		\bottomrule[0.35mm]		
	\end{tabular}
 }
% \vspace{-1em}
\end{table}

\subsection{Performance Comparisons With Benchmarks}

To showcase the role of a certain sensing modality, we assume that all vehicles are equipped with two types of data: received signals and the sensing modality being evaluated. {\color{black}For example, in Table~\ref{sumrate1}, the Pilot-LiDAR scheme indicates that all vehicles are equipped with LiDAR and have access to received pilot signals, while other sensing modalities are assumed to be unavailable.} The sum rate performances are presented in Table~\ref{sumrate1}. The SNR is set as  $\text{SNR}=10\log_{10}(\frac{P}{\sigma^2})=30$dB and pilot sequence length is set to $L_{\rm P}=8$. The Pilot-LiDAR scheme achieves a comparable performance with WMMSE w/ GT CSIT method thanks to the collaborative characterization of communication environment by all on-board LiDARs, demonstrating its greatest role across all sensing modalities. Despite the presence of localization errors and signal disappearing issues, the LoS component information provided by GPS still brings an evident performance enhancement. The role of RGB images falls between LiDAR and GPS, while the performance of the H-MVMM depends on the sensors available in vehicles.

\begin{figure}[!t]
	% \vspace{-0.5em}
	\centering
	\includegraphics[width=1\linewidth]{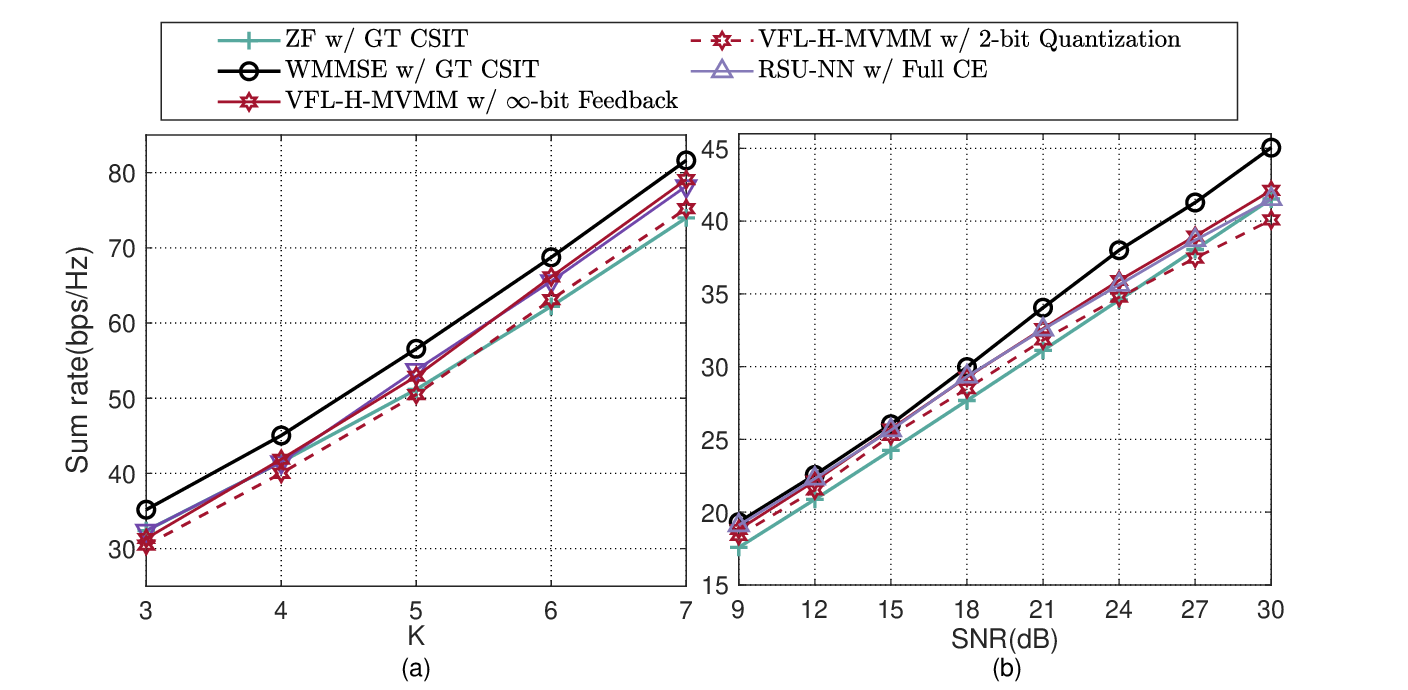}
	\caption{Performances of the offline-trained H-MVMM and benchmarks: (a) vs. $K$ at SNR=$30$dB; (b) vs. SNR at $K=4$.
		\label{rate}}
		% \vspace{-1.5em}
\end{figure}

Fig.~\ref{rate}(a) compares the sum rate achieved by baseline methods and the H-MVMM scheme. The H-MVMM scheme w/ limited feedback can approach the performance of ZF w/ GT CSIT method when $K < 5$ and surpass it as $K$ increases. Furthermore, as $K$ increases, the H-MVMM scheme w/ sufficiently large $B$ approximates the performance of WMMSE w/ GT CSIT method and outperforms RSU-NN w/ full CE method.
Fig.~\ref{rate}(b) depicts the performances of different schemes against different SNRs. We can see that the advantage of the H-MVMM scheme w/ limited feedback is more evident in low SNR regimes. Its performance closely approaches those of WMMSE and RSU-NN. Despite the performance gap between H-MVMM and WMMSE, the H-MVMM scheme enjoys significantly lower complexity. The time complexity of WMMSE in one iteration is $\mathcal{O}(K^2{N^3})$ \cite{WMMSE} while that of offline-trained H-MVMM is on the order of $\mathcal{O}(L_{\rm P}N)$. The H-MVMM scheme achieves a constant increase in time complexity with respect to $N$ and $L_{\rm P}$ ($L_{\rm P} \ll N$).

{\color{black}
\subsection{Robustness Evaluation Under Sensor Imperfections}

In this section, we evaluate the robustness of the proposed scheme against practical sensor imperfections, a critical aspect for deployment in real-world VCN. The experiments are designed to simulate two common, non-ideal scenarios: sensing inaccuracies and occasional sensor faults/blockages. For the first scenario, sensing inaccuracies are emulated by intentionally corrupting the sensory inputs. Specifically, for the LiDAR BEV data $\mathbf{X}_k^{\rm LiDAR}$, we randomly erase point clouds within random angular sectors to simulate object occlusions. For the RGB image modality, we simulate building detection misses and false alarms by applying a random bit-flipping operation to the extracted indicator vector $\mathbf{r}_k$. For the second scenario, occasional sensor faults/blockages are simulated by completely dropping the data from one sensor modality (e.g., LiDAR or RGB) for a short, continuous duration. During this fault period, the system is configured to rely on the latest available historical data from the faulty sensor, fused with the real-time data from other functioning sensors, to assess the proposed scheme's adaptability. These tests, conducted with a user number of $K=4$ across various SNR regimes, aim to validate the inherent robustness and complementary benefits of our multi-modal fusion approach when facing realistic sensor failures.

In the first scenario, we evaluate robustness against sensing inaccuracies by injecting noise into the sensory inputs. Four severity levels are defined by varying the corruption parameters: \textbf{Low} (bit-flip probability $p=0.2$, LiDAR occlusion $30^\circ$), \textbf{Medium} ($p=0.5$, $70^\circ$), and \textbf{High} ($p=0.6$, $100^\circ$). As shown in Fig.~\ref{noise}, under Low noise conditions, the proposed scheme maintains approximately $94\%$ of its ideal performance. Under High noise, it preserves about $91\%$ of the ideal performance. As the environmental noise increases, the performance of the proposed scheme is unavoidably affected. In such cases, the RSU can transmit additional downlink pilots to compensate for the sensing impairment and maintain the target performance.
\begin{figure}[!t]
	\centering
	\includegraphics[width=0.98\linewidth]{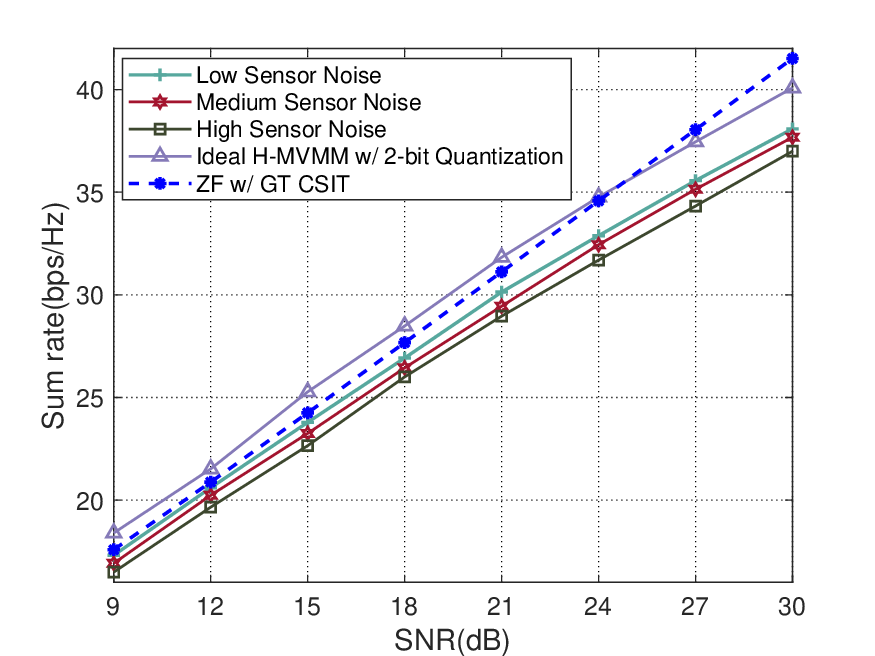}
	\caption{ Performance of the proposed H-MVMM scheme under varying levels of sensing inaccuracies.
	\label{noise}}
\end{figure}

In the second scenario, to evaluate robustness under realistic sensor faults, we simulate four levels of intermittent data blockage. Each level is defined by the fault (frequency × duration): \textbf{Low} (GPS: $1\times[10-20 \text{ slots}]$, LiDAR/RGB: $2\times[20-60 \text{ slots}]$), \textbf{Medium} (GPS: $2\times[10-20 \text{ slots}]$, LiDAR/RGB: $4\times[20-60 \text{ slots}]$), \textbf{High} (GPS: $3\times[10-20 \text{ slots}]$, LiDAR/RGB: $5\times[20-60 \text{ slots}]$), and \textbf{Severe} (GPS: $4\times[10-20 \text{ slots}]$, LiDAR/RGB: $6\times[20-60 \text{ slots}]$). During faults, the system utilizes the latest available historical sensor data. The results demonstrate our scheme's robustness against sensor blockages. As shown in Fig.~\ref{blocl}, under low blockage, the proposed scheme retains approximately  $93\%$ of the ideal performance, proving the effectiveness of multi-modal complementarity. Critically, even under severe blockage, the scheme maintains $88\%$-$91\%$ of its ideal performance. While the performances under high and severe blockage levels show a modest gap compared to the ZF w/ GT CSIT method, this result is justifiable given our scheme's substantially lower computational complexity and pilot overhead.

\begin{figure}[!t]
	\centering
	\includegraphics[width=0.98\linewidth]{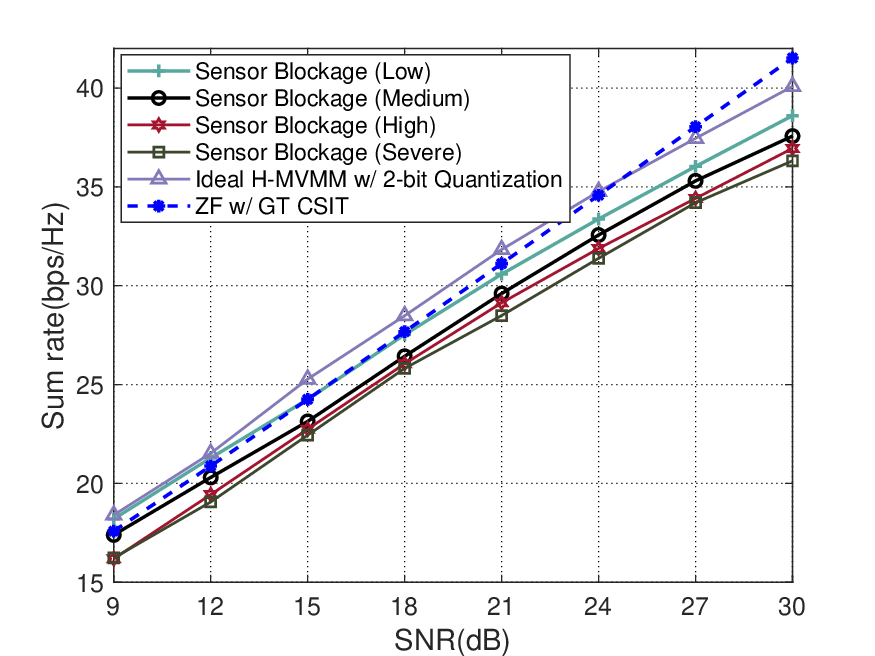}
	\caption{ Performance of the proposed H-MVMM scheme under varying levels of occasional sensor faults/blockages.
	\label{blocl}}
\end{figure}
}

% \vspace{0.5em}
\subsection{Comparisons to CL-Based Schemes}
\label{vsVL}
We present in Table~\ref{setup} the differences in implementing the proposed scheme within VFL and CL frameworks. For clarity, the difference is reflected by the ratio between the dataset volume and the sum rate performance in VFL-based and CL-based schemes. In terms of data consumption required by NN training, VFL-based schemes only require users to transmit a precoding vector of size $N_{\rm t} \times 1$ to RSU in each epoch via UL, thus requiring only a very small amount of data transmission. Specifically, assuming that the real and imaginary parts of a complex number are each represented by $8$ bits, a complex precoding vector occupies $0.25$KB. Thus, the amount of data transmission in VFL is given by $W_{\rm FL} = N_{\rm epoch} \times K \times 0.25$KB, where $N_{\rm epoch}$ is the number of training epochs. {\color{black}For example, in the Pilot-RGB scheme, network training requires $N_{\rm epoch} = 600$ epochs. Hence, the VFL-based scheme incurs a total communication overhead of $600 \times 7 \times 0.25$KB $= 1050$KB $\approx 1.0$MB.} Conversely, CL approaches necessitate gathering training data collected by all users, which pose risks of infringing on user privacy. Furthermore, the enormous data transmission costs render them impractical for real-world applications. {\color{black} In our experiment, the CL-based Pilot-RGB scheme requires a resource usage of $K \times W_{\rm RGB} = 7 \times 14$MB = $98$MB, where $W_{\rm RGB}$ represents the size of locally processed RGB information $\mathbf{x}_k^{\rm RGB}$. Therefore, the communication cost of the VFL-based scheme is only $1.0$MB$/98$MB $ \approx 1.02\%$ of that of the CL-based scheme.}
% The required data transmission amount is the sum of the local data volume of all users.

Despite the need to collect vast amounts of training data, the advantage of CL is that RSU can directly obtain the precoding matrix and avert the quantization errors caused by limited feedback. Nonetheless, as shown in Table~\ref{setup}, the performance losses caused by quantization error are minimal, at less than $6\%$. Therefore, it can be concluded that VFL is a more suitable learning framework for the considered scenario.

\begin{table}[!t]
	\setlength{\abovecaptionskip}{0.1cm} 
	\renewcommand\arraystretch{0.05} 
	\centering
	\caption{Comparison of different schemes under VFL and CL frameworks when $K$=7. The number before the slash represents the commmunication cost required for the VFL-based schemes, while the number after the slash corresponds to the CL-based schemes.}
	\resizebox{1\linewidth}{!}{
		\label{setup}
		\begin{tabular}{@{\hspace{0.6em}}c@{\hspace{0.6em}}|@{\hspace{1em}}c@{\hspace{1em}}|@{\hspace{1em}}c@{\hspace{1em}}}
			\toprule[0.35mm]
			{\textbf{Scheme} }  & {{\textbf{Communication cost}} (MB)} & {{\textbf{Sum rate}} (bps/Hz)}  \\
			% \midrule[0.15mm]
			% \makecell[c]{Uni-Pilot scheme} & \underline{$1.8$}/$84.0$	& \underline{$70.5$}/$74.1$ \\ 
			\midrule[0.15mm]
			\makecell[c]{Pilot-LiDAR} & 
            % $0.08\%$ & $95.8\%$ \\ 
            \underline{$0.4$}/$532.0$ $(0.08\%)$ & \underline{$80.2$}/$83.7$ $(95.8\%)$ \\ 
			\midrule[0.15mm]			
			\makecell[c]{Pilot-RGB} & 
            % $1.12\%$	& $95.4\%$  \\
            \underline{$1.0$}/$98.0$ $(1.02\%)$	& \underline{$72.4$}/$75.9$ $(95.4\%)$ \\
			\midrule[0.15mm]			
			\makecell[c]{Pilot-GPS}  & 
            % $1.65\%$	& $94.2\%$  \\
            \underline{$1.4$}/$85.1$ $(1.65\%)$& \underline{$68.1$}/$72.3$ $(94.2\%)$  \\
			\midrule[0.15mm]		
			\makecell[c]{H-MVMM}& 
            \underline{$1.4$}/$282.8$ $(0.50\%)$ & 
            % $0.50\%$	& $95.1\%$  \\
            \underline{$75.2$}/$79.1$ $(95.1\%)$ \\ 		
			\bottomrule[0.35mm]		
		\end{tabular}	
	}
\end{table}

\subsection{Effectiveness of the Online Model Updating Strategy}
\begin{figure}[!t]
	% \vspace{-0.5em}
	\centering
	\includegraphics[width=1\linewidth]{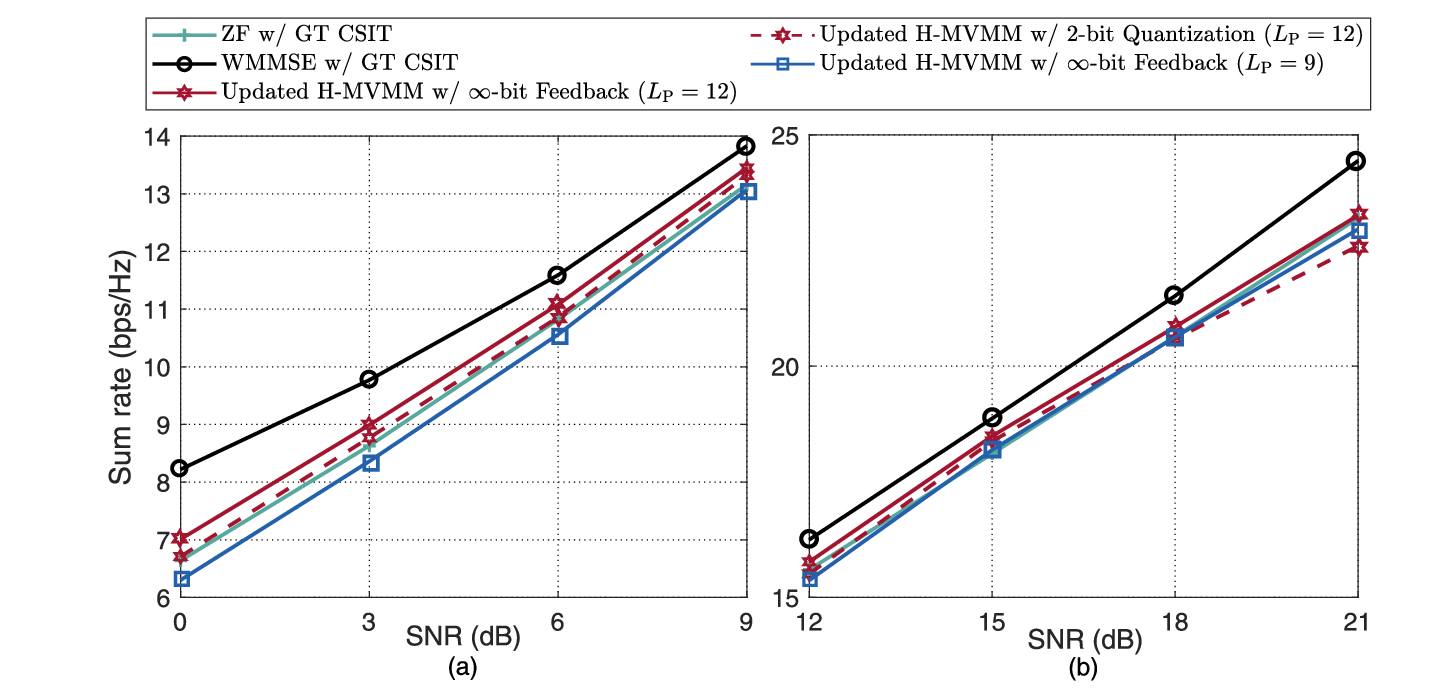}
	\caption{ Sum rate performances of the updated H-MVMM scheme and benchmarks at $K = 3$.
		\label{rateDLCSI}}
	% \vspace{-0.2em}
\end{figure}

% Without loss of generality, in the following experiments, we fix the number of users to $K=3$ to facilitate simulating the performances for multiple schemes under different settings within a relatively short period of time. Note that the numerical results for the configured system parameters are sufficient to serve as a proof of concept for the proposed method.
In this section, we present the performance comparison between the updated H-MVMM scheme after the number of users changes to $K=3$ and the benchmarks. In the following experiments, $(M_1, M_2)$ are configured as $(2,2)$ or $(2,1)$, yielding $L_{\rm P} = K\times M= (12, 9)$.
Fig. \ref{rateDLCSI} illustrates the sum rate performance achieved by the H-MVMM scheme after online model updating using the proposed PCSI-Simulator under different SNRs. It can be observed that, when the pilot sequence length is $L_{\rm P}=12$, the updated H-MVMM scheme w/ $\infty$-bit feedback outperforms the ZF w/ GT CSIT scheme across all SNRs. This demonstrates the accuracy of the surrogate DL CSI labels generated by the PCSI-Simulator. When the pilot sequence length $L_{\rm P}$ decreases to $9$, the updated H-MVMM scheme can still achieve performance close to that of the ZF w/ GT CSIT scheme. Another essential observation is that, at low SNRs (below $12$dB), the performance of the PCSI-Simulator is more sensitive to changes in pilot sequence length. This is because, at low SNRs, the PCSI-Simulator requires more comprehensive DL channel measurements to generate {\color{black}usable} DL CSI labels.
 % and achieves performance close to the WMMSE w/GT CSIT scheme
\begin{table*}[!t]
	\setlength{\abovecaptionskip}{0.04cm} 
	\renewcommand\arraystretch{1} % 进一步缩小行高因子
	\setlength{\aboverulesep}{2.8pt}    % 减少表格线上方间距
	\setlength{\belowrulesep}{1.4pt}   % 减少表格线下方间距
	\tiny
	\caption{Ablation Study on the PCSI-Simulator. The performance of different schemes is reflected in the sum rate (bps/Hz) achieved by the updated H-MVMM under $\infty$-bit feedback  when $L_{\rm P}=12$.}
	% Sum rate performance of the H-MVMM scheme updated by different online model update methods when $K=3$.
	\centering
	\label{tab-aba}
	\resizebox{\linewidth}{!}{
		\begin{tabular}{c|c|c|c|c|c|c|c|c}
			\toprule[0.2mm] % 略微减细表格线
			\multirow{2}{*}{\textbf{Schemes}}  & \multicolumn{8}{c}{\textbf{SNR (dB)}} \\[-2.6pt]  
			\cmidrule[0.1mm]{2-9} % 减细分隔线
			&  0 & 3 & 6 & 9 & 12 & 15 & 18 & 21\\ [-2.2pt] 
			% \cmidrule[0.1mm]{2-9} % 减细分隔线
			% & \multicolumn{8}{c}{\textbf{Sum rate (bps/Hz)}}   \\ [-2.2pt] 
			\midrule[0.1mm]   
			{PCSI-Simulator} & \underline{7.01} & \underline{8.99} & \underline{11.08} & \underline{13.45} & \underline{15.77} & \underline{18.49} & \underline{20.85} & \underline{23.26} \\[-2.8pt] % 添加负间距
			\midrule[0.1mm] 
			\makecell[c]{PCSI-Simulator w/o Codeword-Teacher} & 4.97 & 7.12 & 9.97 & 12.48 & 15.21 & 17.84 & 20.50 & 22.50 \\[-3pt]
			\midrule[0.1mm]
			\makecell[c]{PCSI-Simulator w/o Main-CSI-Net \& Resi-CSI-Net} & 6.19 & 8.13 & 10.02 & 11.99 & 15.20 & 18.31 & 20.77 & 23.15 \\[-3pt]
			\midrule[0.1mm]
			\makecell[c]{UCAL Scheme} & 3.60 & 5.24 & 6.87 & 9.17 & 11.91 & 14.95 & 17.82 & 20.37 \\[-3pt]
			\midrule[0.1mm]
			\makecell[c]{ZF w/GT CSIT} & 6.64 & 8.63 & 10.82 & 13.16 & 15.60 & 18.11 & 20.64 & 23.20 \\[-3pt]
			\midrule[0.1mm]
			\makecell[c]{WMMSE w/GT CSIT} & 8.22 & 9.77 & 11.52 & 13.83 & 16.25 & 18.87 & 21.52 & 24.44 \\[-3pt]
			\bottomrule[0.2mm]
		\end{tabular}
	}
    % \vspace{-1.0em}
\end{table*}

 \begin{figure}[!t]
	\centering
	\includegraphics[width=1\linewidth]{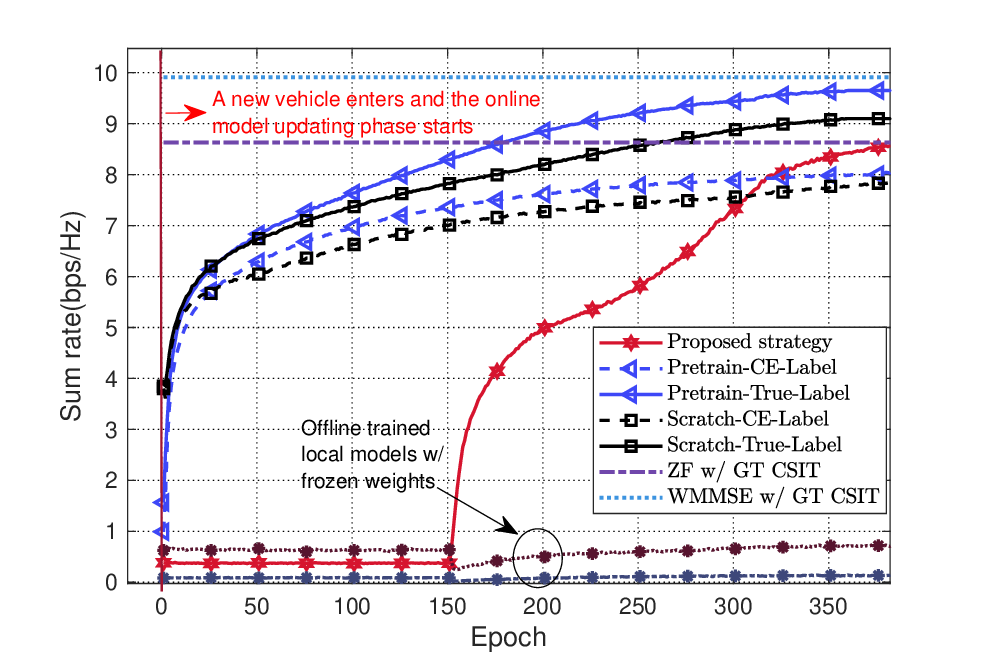}
	\caption{ Sum rate achieved by the updated H-MVMM scheme using different online updating strategies versus number of epochs under $\infty$-bit feedback.
		\label{rateonline}}
	% \vspace{-1.0em}
\end{figure}

\subsection{Adaptability to Changing Users}
In this experiment, we present the adaptability of the H-MVMM scheme to variations in the number of users under different online model updating strategies. Specifically, the user number $K$ changes from $2$ to $3$ during the online deployment phase when SNR is $3$dB. In Fig.~\ref{rateonline}, we first observe that the local models of vehicle 1 and vehicle 2 with frozen weights completely fail when a new vehicle joins, emphasizing the critical need for online weights updates. The schemes updated from pre-trained models exhibit accelerated convergence speed and superior performance compared to schemes trained from scratch. Under this premise, the proposed strategy further exhibits superior performance compared to the two CE label based schemes. This advantage arises from the PCSI-Simulator’s directional pilot transmission mechanism and the denoising effects of Main-CSI-Net and Resi-CSINet. The two pre-trained model-based schemes serve only as theoretical benchmarks since online updates using true CSI labels or complete CE results are impractical in real-world implementations. Our label-free design, despite underperforming these two benchmarks due to inherent label errors, offers adaptability to dynamic user and sensor configurations, ensuring greater practicality.

%\vspace{-0.7em}
\subsection{Ablation Study on PCSI-Simulator}
We conduct ablation experiments to verify the effectiveness of the NN modules included in the PCSI-Simulator. The experiment results are outlined in Table~\ref{tab-aba}.
We observe that without supervision from the Codeword-Teacher, the PCSI-Simulator's performance significantly deteriorates, particularly at low SNRs. This is because, at low SNRs, the DL training codewords selected in the first stage are likely to deviate significantly from the main paths of DL channel, resulting in large errors in the Main-CSI-Net output $\hat{\mathbf{h}}_k^{\rm I}$. Consequently,  without the introduction of pseudo labels, the Resi-CW-Net struggles to learn effective DL training codewords for the second stage through end-to-end learning from the erroneous $\hat{\mathbf{h}}_k^{\rm I}$.  We further validate the roles of Main-CSI-Net and Resi-CSI-Net. As SNR decreases, their contributions become increasingly pronounced. These modules effectively mitigate the noise amplification issue inherent in LS estimator under low-SNR regimes, achieving significant denoising effects through supervision by the online loss function. Finally, considering the existence of partial reciprocity, we propose to directly utilize UL CE as labels to update the H-MVMM. This approach is referred to as UL-CE-As-Label (UCAL). As observed in Table~\ref{tab-aba}, the UCAL scheme exhibits poor performance and shows a significant performance gap compared to the ZF w/ GT CSIT scheme.
% Therefore,, it becomes challenging to select effective DL training codewords for the DL-CSI Generator to generate usable DL CSI.
 % \vspace{-0.5em}
\subsection{Communication Overhead Associated With Model Updates}
\label{overhead}
{\color{black}
The communication overhead for model updating stems from building the two auxiliary datasets, $D_1$ and $D_2$. The collection of dataset $D_1$ requires each user to feedback $N_{\rm c} = 120$ pairs of data, each consisting of a CSI measurement and its corresponding optimal codeword indices label. As outlined in Section \ref{vsVL}, each CSI measurement is a complex-valued vector of length $N=128$, occupying $0.25$KB. Each beam index label is a set of $5$ integer indices. Assuming each index is represented by a $4$-byte ($32$-bit) integer, the label for one sample occupies $5 \times 4 = 20$ bytes. Therefore, the total data volume per user for dataset $\mathcal{D}_1$ is $120 \times (0.25 \text{ KB} + 0.02 \text{ KB}) \approx 32.4 \text{ KB}$. With a typical CSI reporting periodicity of $10$ms \cite{3GPP}, the total time required for this feedback is $120 \times 10 \text{ms} = 1.2\text{s}$.}

 The collection of dataset $\mathcal{D}_2$ involves each user feeding back $N_g=1000$ data samples. Each sample consists of $M=12$ received pilot signals. Quantizing the real and imaginary parts of each signal using 8 bits results in a total of $12 \times 2 \times 8 = 192$ bits (24 bytes) per sample. The total data size per user for $\mathcal{D}_2$ is thus $1000 \times 24 \text{ bytes} \approx 23.4 \text{ KB}$. According to typical 5G NR uplink configurations \cite{3GPP}, one feedback instance can be transmitted within a single uplink slot (e.g., $1$ms for $15$kHz subcarrier spacing). Hence, the total duration for the $1000$ feedback instances is $1000 \times 1 \text{ ms} = 1 \text{ s}$.
 
% In the collection procedure of dataset $\mathcal{D}_2$, each vehicle returns $M=12$ received signals per feedback instance, requiring a total of $192$ bits. With a total number of $1000$ data samples, the overall feedback data size per vehicle is approximately $23.4$KB. According to typical 5G NR uplink configurations defined by 3GPP \cite{3GPP}, one feedback instance can be transmitted within a single uplink slot (e.g., $1$ms for $15$kHz subcarrier spacing). Thus, the total duration for the $1000$ feedback instances is approximately $1$s. 

% (complex IQ values) 
% \vspace{-1.5em}
\section{Conclusions}
\label{conclusion}
This paper proposed a VFL-based precoding approach, termed H-MVMM, for FDD systems. By employing carefully designed data preprocessing methods and a tailored VFL training procedure, the heterogeneous multi-vehicle, multi-modal sensing effectively enhanced the function of pilot transmission. Additionally, a label-free model online updating strategy was introduced, allowing the PCSI-Simulator to generate pseudo CSI labels, which enabled the H-MVMM scheme to update its weights in real time. Simulation results demonstrated that this approach, utilizing significantly shorter pilot sequences, achieved performance closely comparable to classical optimization schemes under perfect CSI.  Furthermore, the H-MVMM was capable of dynamically adapting to changes in user numbers and maintaining strong performance after model updates through the proposed strategy.

To further enhance the H-MVMM scheme's performance and reduce communication overhead associated with the online model update phase, we propose the following two potential directions: {\color{black}i) Address the impact of wireless channel dynamics and unreliability on federated learning to improve training robustness and efficiency.} ii) Develop dynamic user grouping method to effectively cluster users with similar channel characteristics, enabling them to share identical downlink pilots.
% \vspace{-1.5em}

\appendix
\subsection{Proof of Lemma 1}
\label{P1}
The minimizer of the online loss function $\mathcal{L}_{\rm h}$ can be expressed in closed form as follows:
\begin{equation}
	\hat{\mathbf{h}}^* = \mathop{\arg\min}\limits_{\hat{\mathbf{h}}}\frac{1}{M}\Vert \mathbf{y}^{\rm o}-\hat{\mathbf{h}}^{\rm H}\mathbf{S}\Vert_1 = \mathbf{y}^{\rm o}\mathbf{S}^{\dagger}= \mathbf{y}^{\rm o}(\mathbf{S}^{\rm H}\mathbf{S})^{-1}\mathbf{S}^{\rm H},
\end{equation}
yielding $\Vert \hat{\mathbf{h}}^* - {\mathbf{h}} \Vert_{\rm F}^2=\Vert  \mathbf{y}^{\rm o}\mathbf{S}^{\dagger} - {\mathbf{h}} \Vert_{\rm F}^2=\Vert  \mathbf{n}\mathbf{S}^{\dagger} \Vert_{\rm F}^2=\Vert(\mathbf{S}^{\dagger})^{\rm H}\mathbf{n}^{\rm H} \Vert_{\rm F}^2$.
According to the Theorem 3.1 of \cite{prove0}, for any $\alpha>1$, we have:
\begin{equation}
	\label{label-lemma1-2}
	\begin{aligned}
		\text{Pr}\Biggl(&\frac{1}{N}\Vert(\mathbf{S}^{\dagger})^{\mathrm{H}}\mathbf{n}^{\mathrm{H}} \Vert_{\mathrm{F}}^2 > \alpha^2\sigma^2\Vert(\mathbf{S}^{\dagger})^{\mathrm{H}}\Vert^2_{\mathrm{F}} \Biggr) 
		\\
		&\leq \exp\left( -\frac{N}{2}\frac{\Vert(\mathbf{S}^{\dagger})^{\mathrm{H}}\Vert^2_{\mathrm{F}}}{\Vert(\mathbf{S}^{\dagger})^{\mathrm{H}}\Vert^2_{2}}(\alpha-1)^2 \right)
	\end{aligned}
\end{equation}
Given that the orthogonal pilots $\mathbf{S}^{\rm H}\mathbf{S}=P\mathbf{I}_{M}$, we have: $\Vert(\mathbf{S}^{\dagger})^{\mathrm{H}}\Vert^2_{\mathrm{F}}=\Vert (\mathbf{S}^{\rm H}\mathbf{S})^{-1}\mathbf{S}^{\rm H} \Vert^2_{\rm F}= \frac{1}{P^2} \Tr(\mathbf{S}\mathbf{S}^H)=\frac{M}{P}$.
Then, Eq.~\eqref{label-lemma1-2} can be rewritten as
% substituting Eq.~\eqref{label-lemma1-3} into Eq.~\eqref{label-lemma1-2}, 
:
\begin{equation}
	\label{label-lemma1-4}
	\text{Pr}\Biggl(\Vert(\mathbf{S}^{\dagger})^{\mathrm{H}}\mathbf{n}^{\mathrm{H}} \Vert_{\mathrm{F}}^2 > \frac{NM\alpha^2\sigma^2}{P} \Biggr) 
	\leq \exp\left( -\frac{N}{2}(\alpha-1)^2 \right).
\end{equation}
Note that we substitute $\frac{\Vert(\mathbf{S}^{\dagger})^{\mathrm{H}}\Vert^2_{\mathrm{F}}}{\Vert(\mathbf{S}^{\dagger})^{\mathrm{H}}\Vert^2_{2}}$ with its lower bound of $1$.

\vspace{-1em}
\subsection{Proof of Lemma 2}
\label{P2}
Given the continuity of $\eta$ and $\varphi$ in Eq.~\eqref{mapping}, the composite mapping $\varphi^{\prime}:\tilde{{\mathbf{h}}}^{\text{vec}} \mapsto \mathbf{h}^{\text{vec}} $ constitutes a continuous operator. Consequently, each component $ \mathbf{h}^{\text{vec}}_n = \varphi^{\prime}(\tilde{{\mathbf{h}}}_n^{\text{vec}}),n=1,2,\cdots,2N$, preserves continuity over the compact set $\mathcal{C}$. Since $\varphi^{\prime}$ is continuous on $\mathcal{C}$, there exists $a > 0$ such that $|\varphi^{\prime}(\tilde{{\mathbf{h}}}^{\text{vec}})| \leq a, \forall \tilde{{\mathbf{h}}}^{\text{vec}} \in \mathcal{C}$. This boundedness ensures $q$-integrability:
\begin{equation}
	\int_{\Theta} |\varphi^{\prime}(\tilde{{\mathbf{h}}}^{\text{vec}})|^q \, d(\tilde{{\mathbf{h}}}^{\text{vec}}) \leq a^q \mathrm{vol}(\mathcal{C}) < \infty, \quad \forall q > 0.
\end{equation}
Then, as stated in Theorem 9.5.3 of \cite{prove2}, for any $\varepsilon_n > 0$ and $q \geq 1$, there exists a ReLU DNN $\mathcal{G}^n_{\text{DNN}}(\tilde{{\mathbf{h}}}^{\text{vec}})$ with a large enough width $W_n$ and depth $D_n<2(\lfloor\log_2(2N)\rfloor)$, satisfying: $
\Vert \mathcal{G}^n_{\text{DNN}}(\tilde{{\mathbf{h}}}^{\text{vec}}) - \varphi^{\prime}(\tilde{{\mathbf{h}}}^{\text{vec}}) \Vert_{q} < \varepsilon_n^{\frac{q}{2}}$.
Taking the $\frac{1}{q}$-power norm yields: $\Vert \mathcal{G}^n_{\text{DNN}}(\tilde{{\mathbf{h}}}^{\text{vec}}) - \varphi^{\prime}(\tilde{{\mathbf{h}}}^{\text{vec}}) \Vert_{q}^{\frac{1}{q}} < \varepsilon_n^{\frac{1}{2}}$.
Since both $\mathcal{G}^n_{\text{DNN}}(\tilde{{\mathbf{h}}}^{\text{vec}})$ and $\varphi^{\prime}(\tilde{{\mathbf{h}}}^{\text{vec}})$ are continuous over the compact set $\mathcal{C}$, according to Proposition 2.2 of \cite{prove3}, we have:
\begin{equation}
	\begin{aligned}
		&\sup | \mathcal{G}^n_{\text{DNN}}(\tilde{{\mathbf{h}}}^{\text{vec}}) - \varphi^{\prime}(\tilde{{\mathbf{h}}}^{\text{vec}})| \\ &= \lim_{q \to \infty} | \mathcal{G}^n_{\text{DNN}}(\tilde{{\mathbf{h}}}^{\text{vec}}) - \varphi^{\prime}(\tilde{{\mathbf{h}}}^{\text{vec}})|_q^{\frac{1}{q}}< \varepsilon_n^{\frac{1}{2}}.
	\end{aligned}
\end{equation}
By composing all the $2N$ component DNNs $\{\mathcal{G}^n_{\text{DNN}}\}_{n=1}^{2N}$ in parallel, a larger DNN is constructed with width $W=\sum_{n=1}^{2N}W_n$ and preserved depth $D=D_n<2(\lfloor\log_2(2N)\rfloor)$, such that:
\begin{equation}
	\Vert \mathcal{G}^n_{\text{DNN}}(\tilde{{\mathbf{h}}}^{\text{vec}}) - \varphi^{\prime}(\tilde{{\mathbf{h}}}^{\text{vec}}) \Vert^2_2 \leq \varepsilon=\sum_{n=1}^{2N}\varepsilon_n
\end{equation}
\vspace{-2.5em}
\subsection{Proof of Lemma 3}
\label{P3}
We first prove that $\hat{\mathbf{h}}$ lies within some compact sets. Using the traiangular inequality, we have: $\Vert \hat{\mathbf{h}}\Vert _2^2 \leq \Vert {\mathbf{h}}\Vert_2^2 + \Vert \hat{\mathbf{h}} - {\mathbf{h}}\Vert_2^2$. Given the physically realistic Saleh-Valenzuela channel model adopted in the SynthSoM dataset, the channel $\mathbf{h}$ has finite energy. Hence, there exists a finite constant $C_{\rm h}$, such that: $\|\mathbf{h}\|^2_2 \le C_{\rm h}$. According to the consistency property of the online loss function, we obtain:
\begin{equation}
	\Vert \hat{\mathbf{h}}\Vert _2^2 \leq \Vert {\mathbf{h}}\Vert_2^2 + \Vert \hat{\mathbf{h}} - {\mathbf{h}}\Vert_2^2 \leq C_{\rm h} + \frac{NM\alpha^2\sigma^2}{P} .
\end{equation}
Consequently, $\hat{\mathbf{h}}$ is within the compact set $\hat{\mathcal{H}}=\{\hat{\mathbf{h}}:\Vert \hat{\mathbf{h}}\Vert_2^2 \leq C_{\rm h} + \frac{NM\alpha^2\sigma^2}{P}\}$ with probability of at least $1-\exp\left( -\frac{N}{2}(\alpha-1)^2 \right)$ for any $\alpha>1$.

Since the shape of $\hat{\mathcal{H}}$ is not affected by $\mathbf{y}^{\rm o}$ and $\mathbf{S}$, the compact set $\hat{\mathcal{H}}$ is trivial. Given the continuity of the loss function $\mathcal{L}_{\rm h}$ in  $\hat{\mathbf{h}}$ and $\{\mathbf{y}^{\rm o},\mathbf{S}\}$, the Maximum Theory in Theorem 9.14 of \cite{prove4} guarantees that the solution correspondence:
\begin{equation}
	\mathcal{R}^*(\hat{\mathbf{h}}) = \mathop{\arg\min}\limits_{\hat{\mathbf{h}}}\frac{1}{M}\Vert \mathbf{y}^{\rm o}-\hat{\mathbf{h}}^{\rm H}\mathbf{S}\Vert_1
\end{equation}
is upper semi-continuous (usc) on $\mathcal{C}$. When the only point in $\mathcal{R}^*(\hat{\mathbf{h}})$ is the minimizer $\hat{\mathbf{h}}^*$ in Eq.~\eqref{mapping}, 
$\mathcal{R}^*(\cdot)$ is a single-valued correspondence. By Theorem 9.12 of \cite{prove4}, a single-valued usc correspondence constitutes a continuous function. Hence, $\mathcal{R}^*(\hat{\mathbf{h}})=\varphi(\tilde{\mathbf{h}})$ is continuous over $\mathcal{C}$.
\vspace{-1em}
\subsection{Proof of Theorem 1}
\label{P4}
We first demonstrate that the input of DNN, i.e., LS estimates $\tilde{\mathbf{h}}$, lies within a compact set. Without loss of generality, in this subsection, we unify the pilot matrices used in two DL training phases as $\mathbf{S} \in \mathbb{C}^{N \times M}$, and omit the subscript $k$. Firstly, substituting the received signals expression in Eq.~\eqref{y} into the LS estimator, we obtain:
\begin{equation}
	\tilde{\mathbf{h}} = \mathbf{y}\mathbf{S}^{\dagger}
	= (\mathbf{h}^{\rm H}\mathbf{S} + \mathbf{n})\mathbf{S}^{\dagger}
	= \mathbf{h}^{\rm H}\mathbf{S}(\mathbf{S}^{\rm H}\mathbf{S})^{-1}\mathbf{S}^{\rm H} + \mathbf{n}\mathbf{S}^{\dagger}.
\end{equation}
where $\mathbf{h}$ is the ground truth channel and $\mathbf{n}\sim \mathcal{CN}(\mathbf{0},\sigma^2\mathbf{I}_{M})$ is the additive Gaussian noise. Since the matrix $\mathbf{S}(\mathbf{S}^{\rm H}\mathbf{S})^{-1}\mathbf{S}^{\rm H}$ is a projection onto the column space of $\mathbf{S} \in \mathbb{C}^{N\times M}$, we denote this projection matrix by: $\mathbf{P}_{\rm S} = \mathbf{S}(\mathbf{S}^{\rm H}\mathbf{S})^{-1}\mathbf{S}^{\rm H}$. Thus, the LS estimates can be simplified as: $\tilde{\mathbf{h}} = \mathbf{h}^{\rm H}\mathbf{P}_{\rm S} + \mathbf{n}\mathbf{S}^{\dagger}$.

Then, using the triangle inequality, we have:
\begin{equation}
	\label{key1}
	\|\tilde{\mathbf{h}}^{\rm vec}\|_2^2 = \|\tilde{\mathbf{h}}\|_2^2 = \|\mathbf{h}^{\rm H}\mathbf{P}_{\rm S} + \mathbf{n}\mathbf{S}^{\dagger}\|^2_2
	\le \|\mathbf{h}^{\rm H}\mathbf{P}_{\rm S}\|^2_2 + \|\mathbf{n}\mathbf{S}^{\dagger}\|^2_2.
\end{equation}
Since $\Vert \mathbf{P}_{\rm S} \Vert^2_2 = \Vert \frac{1}{P}\mathbf{S}\mathbf{S}^{\rm H}\Vert^2_2=\Vert \frac{1}{P}\mathbf{S}^{\rm H}\mathbf{S}\Vert^2_2=1$, then we have: $\|\mathbf{h}^{\rm H}\mathbf{P}_{\rm S}\|^2_2 \le \|\mathbf{h}^{\rm H}\|^2_2 \le C_{\rm h}$.

As $\|\mathbf{n}\mathbf{S}^{\dagger}\|^2_2=\mathbf{n}\mathbf{S}^{\dagger}(\mathbf{S}^{\dagger})^{\rm H}\mathbf{n}^{\rm H}$, $\|\mathbf{n}\mathbf{S}^{\dagger}\|^2_2$ is a quadratic form of a Gaussian vector. Clearly, we have: $\|\mathbf{n}\mathbf{S}^{\dagger}\|^2_2 \leq \|\mathbf{n}\|^2_2 \|\mathbf{S}^{\dagger}\|^2_2 = \frac{1}{P}\| \mathbf{n}\|^2_2 $. Since $\mathbf{n}  \sim \mathcal{CN}(\mathbf{0},\sigma^2\mathbf{I}_M)$, its squared norm divided by $\sigma^2$ follows a Chi-squared distribution with $2M$ degrees of freedom, i.e., $\frac{1}{\sigma^2}\|\mathbf{n}\|^2_2\sim \chi^2_{2M}$. Then, for any small probability $\epsilon > 0$, we have:
\begin{equation}
	\text{Pr}\Bigl(  \|\mathbf{n}\|^2_2 \leq {\sigma^2}Q_{\chi^2_{2M}}(1-\epsilon) \Bigr)=1-\epsilon,
\end{equation}
where $Q_{\chi^2_{2M}}(1-\epsilon)$ is the $(1-\epsilon)$-quantile of the chi-squred distribution with $2M$ degrees of freedom. Consequently, for any chosen small probability $\epsilon>0$, the  inequality $\|\mathbf{n}\mathbf{S}^{\dagger}\|^2_2 \leq \frac{1}{P}\|\mathbf{n}\|^2_2 \leq \frac{\sigma^2}{P}Q_{\chi^2_{2M}}(1-\epsilon)$ holds
with probability at least $1-\epsilon$.

Revisiting Eq.~\ref{key1}, we have:
\begin{equation}
	\|\tilde{\mathbf{h}}^{\rm vec}\|_2^2 \le \|\mathbf{h}^{\rm H}\mathbf{P}_{\rm S}\|^2_2 + \|\mathbf{n}\mathbf{S}^{\dagger}\|^2_2 
	\leq C_{\rm h} + \frac{\sigma^2}{P}Q_{\chi^2_{2M}}(1-\epsilon),
\end{equation}
with probability at least $1-\epsilon$. As such, $\tilde{\mathbf{h}}$ will be in the compact set given by $\mathcal{C} = \{\tilde{\mathbf{h}}^{\rm vec} : \|\tilde{\mathbf{h}}^{\rm vec}\|_2^2 \le C_{\rm h} + \frac{\sigma^2}{P}Q_{\chi^2_{2M}}(1-\epsilon)\}$ w.h.p. at least $1-\epsilon$.

Then, given that $\hat{\mathbf{h}}^*_{\text{vec}}=\eta(\hat{\mathbf{h}}^*)\in \mathbb{R}^{2N \times 1}$, the triangle inequality is then applied to derive:
\begin{equation}
	\begin{aligned}
		&\sup_{\tilde{\mathbf{h}}^{\rm vec}\in \mathcal{C}} \| \eta^{-1}(\mathcal{G}_{\text{DNN}}(\tilde{{\mathbf{h}}}^{\text{vec}}))- \mathbf{h}\|^2_2 \\
		&\leq \sup_{\tilde{\mathbf{h}}^{\rm vec}\in \mathcal{C}} \| \eta^{-1}(\mathcal{G}_{\text{DNN}}(\tilde{{\mathbf{h}}}^{\text{vec}}))- \hat{\mathbf{h}}^* \|^2_2 + \|  \hat{\mathbf{h}}^* - \mathbf{h}\|_2^2 \\
		&\leq \sup_{\tilde{\mathbf{h}}^{\rm vec}\in \mathcal{C}} \| \mathcal{G}_{\text{DNN}}(\tilde{{\mathbf{h}}}^{\text{vec}})- \hat{\mathbf{h}}^*_{\text{vec}} \|^2_2 + \|  \hat{\mathbf{h}}^* - \mathbf{h}\|_2^2 \\
		&\leq  \varepsilon + \frac{NM\alpha^2\sigma^2}{P}.
	\end{aligned}
\end{equation}
In the inequality, according to Lemma 2, a DNN with bounded number
of layers can universally approximate $\varphi$ with arbitrarily small error such that $\sup_{\tilde{\mathbf{h}}^{\rm vec}\in \mathcal{C}} \| \mathcal{G}_{\text{DNN}}(\tilde{{\mathbf{h}}}^{\text{vec}})- \hat{\mathbf{h}}^*_{\text{vec}} \|^2_2 \leq \varepsilon$ given that the inverse mapping $\varphi$ is continuous. The second term in the last inequality is owed to the consistency property of the online loss function.

	\begin{IEEEbiography}[{\includegraphics[width=1in,height=1.25in,clip,keepaspectratio]{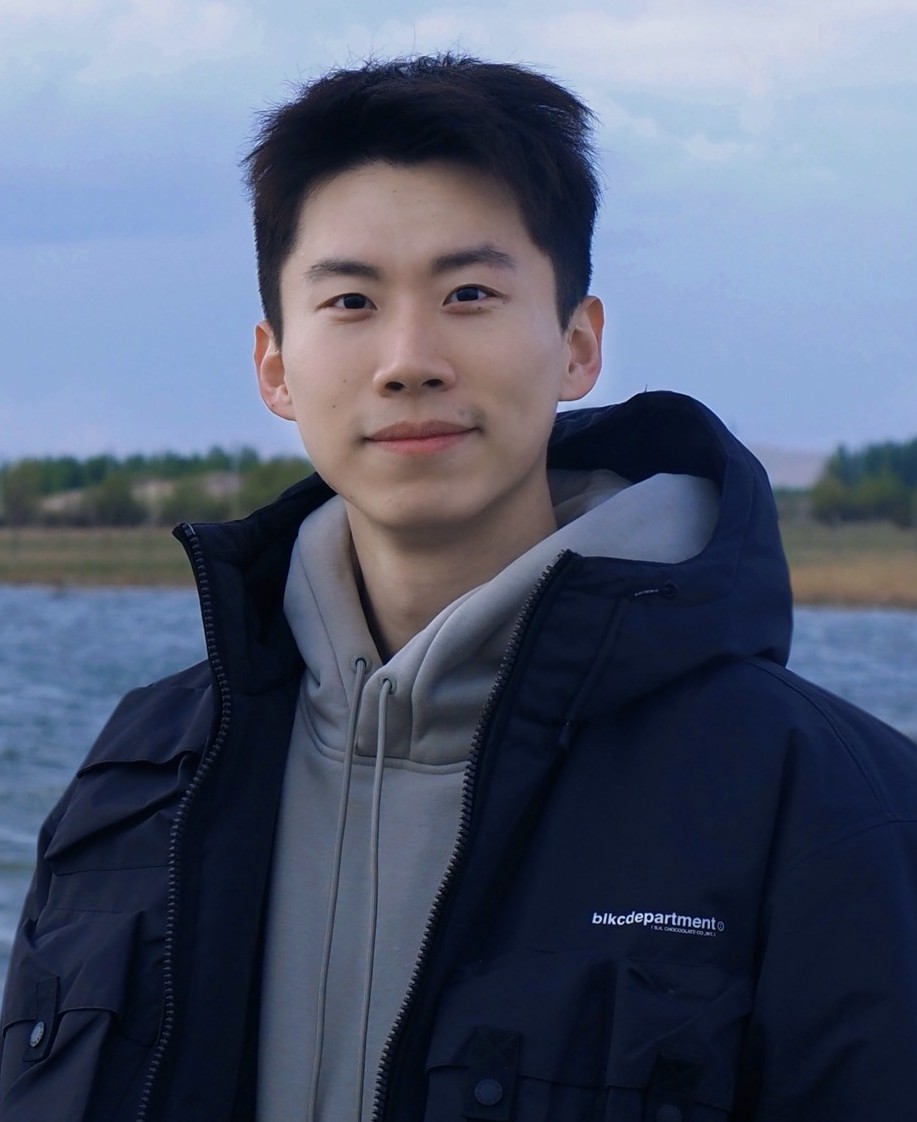}}]{Haotian Zhang}(Graduate Student Member, IEEE)
	received the B.E. degree in network engineering from University of Electronic Science and Technology of China, Chengdu, China, in 2022. He is currently pursuing the Ph.D. degree with the School of Electronics, Peking University, Beijing, China. His current research interest is multi-modal sensing assisted communication system design.\end{IEEEbiography}
 		
	\begin{IEEEbiography}[{\includegraphics[width=1in,height=1.25in,clip,keepaspectratio]{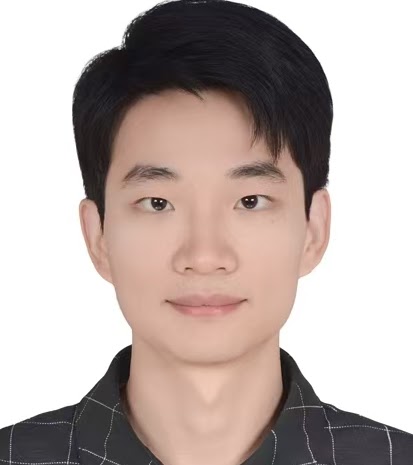}}]{Shijian Gao}(Member, IEEE) received the Ph.D. degree from the University of Minnesota, Minneapolis, MN, USA, in 2022. After graduation, he was a Senior RF Engineer at the Samsung SoC Laboratory, San Diego, CA, USA. In February 2024, he joined the Internet of Things Thrust, The Hong Kong University of Science and Technology (Guangzhou), Guangzhou, China, as an Assistant Professor. His research interests include statistical signal processing, AI for wireless,  and low-altitude networking systems. He is a co-recipient of the 2021 MICCAI Young Scientist Paper Award, the 2024 China Communications Society Science and Technology Paper Award, and the 2025 IEEE WCSP Best paper award. He serves as an Associate Editor for IET Communications and IEEE Communication Letters.
	\end{IEEEbiography}

    \begin{IEEEbiography}[{\includegraphics[width=1in,height=1.25in,clip,keepaspectratio]{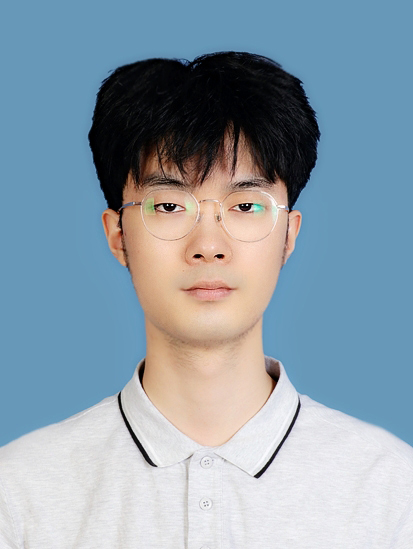}}]{Weibo Wen}(Graduate Student Member, IEEE)
	received the B.E. degree in communication engineering from University of Electronic Science and Technology of China, Chengdu, China, in 2024. He is currently pursuing the Ph.D. degree with the School of Electronics, Peking University, Beijing, China. His research interests include wireless foundation model empowered communication system design.\end{IEEEbiography}
    
	\begin{IEEEbiography}[{\includegraphics[width=1in,height=1.25in,clip,keepaspectratio]{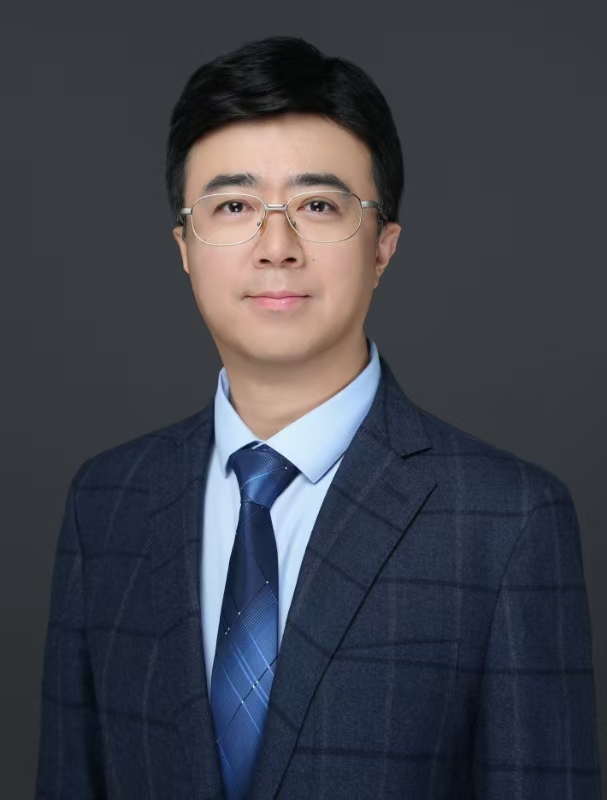}}]{Xiang Cheng} (Fellow, IEEE)  received the joint Ph.D. degree from Heriot-Watt University and The University of Edinburgh, Edinburgh, U.K., in 2009. He is currently a Boya Distinguished Professor with Peking University. His research focuses on the in-depth integration of communication networks and artificial intelligence, including intelligent communication networks and connected intelligence, the subject on which he has published more than 280 journals and conference papers, 11 books, and holds 32 patents. He was a recipient of the IEEE Asia–Pacific Outstanding Young Researcher Award in 2015 and the Xplorer Prize in 2023. He was a co-recipient of the 2016 IEEE JOURNAL ON SELECTED AREAS IN COMMUNICATIONS Best Paper Award: Leonard G. Abraham Prize and the 2021 IET Communications Best Paper Award: Premium Award. He has also received the Best Paper Awards at IEEE ITST’12, ICCC’13, ITSC’14, ICC’16, ICNC’17, GLOBECOM’18, ICCS’18, and ICC’19. He has been a Highly Cited Chinese Researcher since 2020. In 2021 and 2023, he was selected into two world scientist lists, including the World’s Top 2\% Scientists released by Stanford University and top computer science scientists released by Guide2Research. He has served as the symposium lead chair, the co-chair, and a member of the technical program committee for several international conferences. He led the establishment of four Chinese standards (including industry standards and group standards) and participated in the formulation of ten 3GPP international standards and two Chinese industry standards. He is currently a Subject Editor of IET Communications; an Associate Editor of IEEE TRANSACTIONS ON WIRELESS COMMUNICATIONS, IEEE TRANSACTIONS ON INTELLIGENT TRANSPORTATION SYSTEMS, IEEE WIRELESS COMMUNICATIONS LETTERS, and Journal of Communications and Information Networks. He was a Distinguished Lecturer of the IEEE Vehicular Technology Society.\end{IEEEbiography}
	
    \begin{IEEEbiography}[{\includegraphics[width=1in,height=1.25in,clip,keepaspectratio]{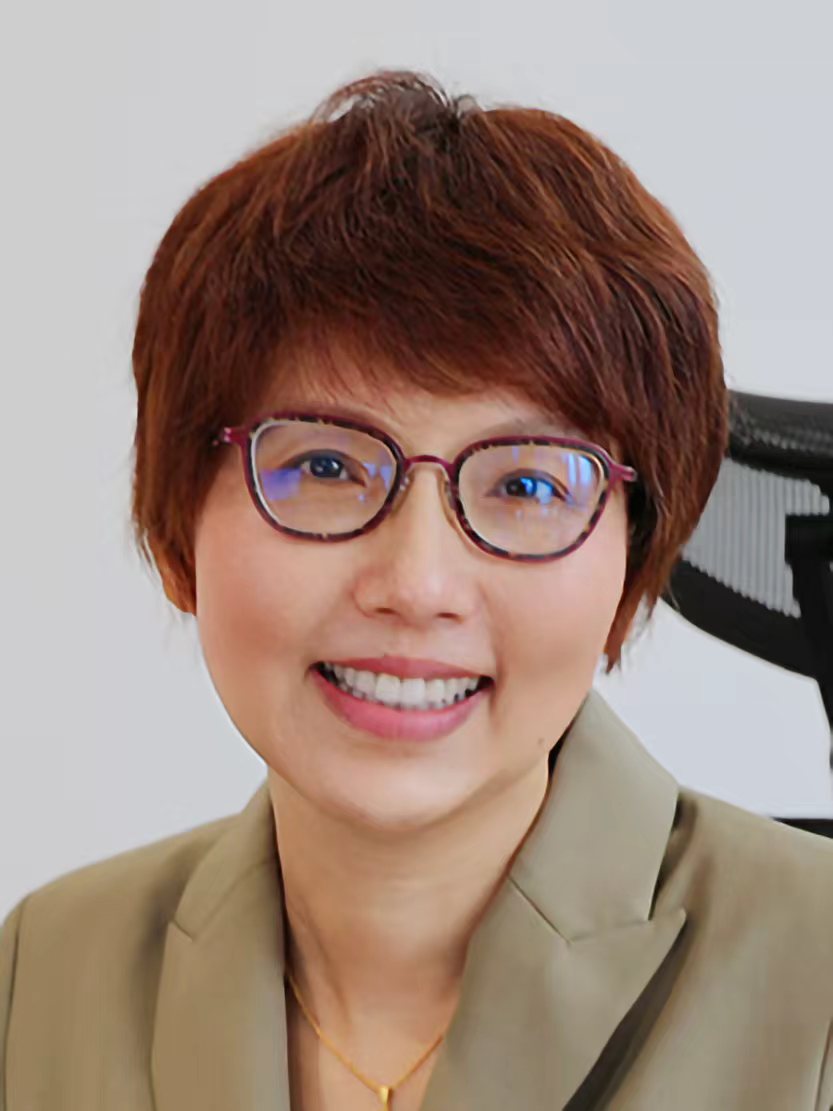}}]{Liuqing Yang}(Fellow, IEEE)
	received the Ph.D. degree from the University of Minnesota, Minneapolis, MN, USA, in 2004. She is a Fellow of IEEE and AAIA. Prof. Yang has been a faculty member with University of Florida, Colorado State University, and University of Minnesota, and is currently a Chair Professor with the Hong Kong University of Science and Technology (Guangzhou), where she serves as the Acting Director of the Low-Altitude Systems and Economy Research Institute (LASERi), and the Head of the Intelligent Transportation (INTR) Thrust. Her research interests include communications, sensing, and networked intelligence, subjects on which she has published more than 400 journal and conference papers, four book chapters, and five books. She is a recipient of the ONR YIP Award in 2007, the NSF CAREER Award in 2009, and multiple Best Paper Awards. Prof. Yang is an Executive Editorial Committee (EEC) Member of the IEEE Transactions on Wireless Communications. She has also served as the Editor-in-Chief of IET Communications, on the editorial board for an array of elite journals including IEEE Transactions on Signal Processing, IEEE Transactions on Communications, and the IEEE Transactions on Intelligent Transportation Systems, in various roles of IEEE ComSoc and IEEE ITSS, as well as in leadership roles for many conferences. \end{IEEEbiography}
    
\end{document}